\documentclass[lettersize,journal]{IEEEtran}
\usepackage{amsmath,amsfonts}
\usepackage{algorithmic}
\usepackage{algorithm}
\usepackage{array}
\usepackage[caption=false,font=normalsize,labelfont=sf,textfont=sf]{subfig}
\usepackage{textcomp}
\usepackage{stfloats}
\usepackage{url}
\usepackage{verbatim}
\usepackage{graphicx}
\usepackage{cite}

\usepackage{microtype}
\usepackage{setspace}
\usepackage[numbers]{natbib}
\usepackage{hyperref}
\newtheorem{definition}{Definition}
\usepackage{booktabs} 
\usepackage{multirow}
\usepackage{tabularx}
\usepackage{listings}
\begin{document}

\title{Efficient and Universal Watermarking for LLM-Generated Code Detection}

\author{Boquan Li, Zirui Fu, Mengdi Zhang, Peixin Zhang, Jun Sun and Xingmei Wang
\thanks{Boquan Li, Zirui Fu, Xingmei Wang are with College of Computer Science and Technology, Harbin Engineering University.
Mengdi Zhang, Peixin Zhang, Jun Sun are with School of Computing and Information Systems, Singapore Management University.}
\thanks{Corresponding to Peixin Zhang at pxzhang@smu.edu.sg and Xingmei Wang at wangxingmei@hrbeu.edu.cn.}
}

\maketitle

\begin{abstract}
Large language models (LLMs) have significantly enhanced the usability of AI-generated code, providing effective assistance to programmers. 
This advancement also raises ethical and legal concerns, such as academic dishonesty and the generation of malicious code.
For accountability, it is imperative to detect whether a piece of code is AI-generated.
Watermarking is broadly considered a promising solution and has been successfully applied to identify LLM-generated text. 
However, existing efforts on code are far from ideal, suffering from limited universality and excessive time and memory consumption.
In this work, we propose a plug-and-play watermarking approach for AI-generated code detection, named \textbf{ACW} (\textbf{A}I \textbf{C}ode \textbf{W}atermarking).
\textbf{ACW} is training-free and works by selectively applying a set of carefully-designed, semantic-preserving and idempotent code transformations to LLM code outputs.
The presence or absence of the transformations serves as implicit watermarks, enabling the detection of AI-generated code.
Our experimental results show that \textbf{ACW} effectively and efficiently detects AI-generated code, preserves code utility, and is resilient against potential code disruptions.
Especially, \textbf{ACW} is universal across different LLMs, addressing the limitations of existing approaches.
\end{abstract}


\section{Introduction}

\IEEEPARstart{L}{arge} language models (LLMs)~\citep{llm-survey} mark a milestone in the progress of artificial intelligence (AI) and prompt a variety of applications~\citep{summary-application}, one of which is code generation.
Multiple LLMs, such as ChatGPT~\citep{openai2023chatgpt} and Qwen2.5-Coder~\citep{hui2024qwen2}, are widely adopted for code generation.
While LLMs improve programmers' productivity significantly, they can be misused as well.
For instance, \citet{malware-generation} showed how an attacker could use ChatGPT to construct ransomware.
In addition, students may dishonestly leverage LLMs to generate programming assignments.
For accountability, it is an urgent need to detect whether a piece of code is AI-generated.

Existing efforts on detecting AI-generated content mostly focus on text~\cite{zhao2025sok} and are broadly categorized into two groups. 
The first group includes passive detectors~\citep{detect-gpt,bao2024fast,hu2023radar,gptzero} which build binary classifiers to distinguish between AI-generated and human-written text.
The second group focuses on watermarking~\citep{watermark-for-llm,giboulot2024watermax,zhaoprovable,liu2024adaptive,christ2024undetectable,zhang2024remark,dathathri2024scalable}, which actively embeds hidden `patterns' (a.k.a.~watermarks) into AI-generated text, and determines whether given text is generated based on the presence of the pattern.
As the state-of-the-art approaches, \citet{watermark-for-llm} propose WLLM (Watermarking LLMs), which splits a vocabulary into ‘green’ and ‘red’ tokens and softly promotes the ‘green’ ones during text generation, thereby a given piece of text can be concluded AI-generated if abundant ‘green’ tokens are present.
\citet{giboulot2024watermax} propose WaterMax, which embeds watermarks by generating multiple candidate text chunks from an LLM and selecting the candidates that maximize a predefined test statistic, and then detects the watermark by evaluating the same statistic on given text.

In comparison to text, detecting AI-generated code based on watermarking remains an emerging task, with state-of-the-art approaches extending from WLLM.
\citet{who-wrote} propose a selective watermarking technique via entropy thresholding named SWEET, which prompts ‘green’ tokens only at positions with high entropy, so as to preserve the utility of AI-generated code.
Further, \citet{kim2025marking} propose a syntax token preserving watermarking approach named STONE, which extends SWEET and further improves the utility of code.
In addition, some other code watermarking efforts~\citep{guan2024codeip,ning2024mcgmark,llm-ip,codemark,yang2024srcmarker} focusing on diverse tasks such as protecting LLM APIs, are not suitable for the task of AI-generated code detection.

However, existing efforts are far from ideal.
First, the state-of-the-art approaches demonstrate limited accuracy in detecting LLM-generated code (as our empirical evidence in Section~\ref{sec:experiment}).
Second, the LLM manipulation-based watermarking strategies present significant time and memory consumption, which slows down LLM code generation.
Third, they are restricted by open-source LLMs, lacking universality, and thus cannot adapt to LLM updates and optimizations.

In response to the above problems, in this work, we propose a plug-and-play code watermarking approach for AI-generated code detection, named \textbf{ACW} (\textbf{A}I \textbf{C}ode \textbf{W}atermarking).
The key ingredient of \textbf{ACW} is an (extensible) set of 46 carefully-designed, semantic-preserving and idempotent code transformations, which are categorized into refactoring, reordering, and formatting groups, and some of them are designed with certain random factors, such as transformation strategies based on the hashcode of operands, to confuse potential adversaries.

For watermark embedding, given an LLM, \textbf{ACW} systematically post-processes the generated code by selectively applying our designed transformations.
For watermark identification, given any code, \textbf{ACW} identifies the existence of watermarks by checking whether the transformations have been applied.
Note that our transformations are designed to be idempotent, enabling the application status of a certain transformation to be easily identified by checking whether applying the transformation repeatedly changes the code or not.
Ultimately, the watermarks serve as indicators for AI-generated code detection.

In this work, we mainly make the following contributions.

\begin{itemize}
    \item We propose \textbf{ACW}, a novel code watermarking approach for AI-generated code detection, which is plug-and-play and universal across different LLMs.
    Moreover, \textbf{ACW} is efficient and works in a training-free manner, without the need to access or modify LLMs.
    
    \item The semantic-preserving transformations in \textbf{ACW} preserve the utility of code after watermarking.
    Moreover, \textbf{ACW} is resilient against potential code disruptions.
    Even if the embedded watermark is partially modified, the remaining transformations still ensure its discriminability.

    \item We systematically evaluate \textbf{ACW} based on Python code and projects generated using multiple LLMs and prompt datasets.
    Experimental results demonstrate that \textbf{ACW} is effective as well as efficient in detecting AI-generated code, achieving accuracy over 97\%, true positive rates over 94\%, false positive rates below 1\%, and watermarking time below 0.1 seconds average on one piece of code, significantly outperforming existing approaches.
    Moreover, \textbf{ACW} preserves code utility, remains resilient against code disruptions, and is universal across different LLMs.
\end{itemize}

The rest of this paper is organized as follows.
We define our research problem in Section~\ref{sec:preliminary} and detail our approach of \textbf{ACW} in Section~\ref{sec:approach}.
Our experiments are presented in Section~\ref{sec:experiment} and a further ablation study is presented in Section~\ref{sec:study}.
We then review related work in Section~\ref{sec:relatedwork} and conclude this work in Section~\ref{sec:conclusion}.
Finally, we release our code, data and an online appendix in Section~\ref{sec:availability}.

\section{Preliminary}
\label{sec:preliminary}

In this section, we define our research problem and analyze why existing approaches fail to solve the problem satisfactorily.

\subsection{Problem Definition} \label{subsec:problem}

There are multiple reasons we would like to know whether a piece of code is AI-generated, such as verifying the originality of code submissions from students or employees, or clarifying copyright ownership between humans and AI.
The high-level research problem we aim to address is: \emph{Given a piece of code, how can we determine if it is generated by AI?}

In this work, we address the problem based on watermarking, which is broadly considered an effective solution.
It works by embedding distinct patterns into AI-generated code, allowing us to conclude whether given code is generated.
Formally, we define a code watermarking approach as follows.

\begin{definition} \label{def:codewatermark} 
	A code watermarking approach consists of a function pair $(E, I)$, where $E$ is a watermark embedding function and $I$ is a watermark identification function.
	Function $E$ takes a code snippet (or project) $\mathcal{C}$ and a watermark $\omega$ as inputs, and produces a watermarked code counterpart, denoted as $\mathcal{C}^\omega$. 
	Function $I$ takes given code as input and examines the presence of watermarks.
\end{definition}

\emph{Requirements.}
We consider a code watermarking approach to address the defined problem satisfactorily, only if it satisfies the following requirements.

\begin{itemize}
	\item
	\emph{Discriminability}: $I(\mathcal{C}^\omega) \neq I(\mathcal{H})$, where $\mathcal{H}$ represents a human-written code snippet (or project).
    This ensures the AI-generated code must be discriminative from human-written code.
    \item 
    \emph{Efficiency}:
    $T(G(\mathcal{C})+E(\mathcal{C})) \approx T(G(\mathcal{C}))$, where $T$ is a time counting function and $G$ is a code generation function.
    This ensures the time for watermarking must be negligible, making little impact on the efficiency of code generation.
	\item
	\emph{Utility}:
    $\mathcal{C} = \mathcal{C^\omega}$ for any code snippet (or project) $\mathcal{C}$ and watermark $\omega$, where $=$ denotes functional equivalence.
    This ensures the watermarked code must maintain the same functionality as the original code before watermarking.
	\item
	\emph{Resilience}:
    $I(\mathcal{A}(\mathcal{C}^\omega)) = I(\mathcal{C}^\omega)$, where $\mathcal{A}$ represents code optimizations or modifications.
    This ensures the detectability of the watermark under potential code disruptions.
\end{itemize}

\subsection{Existing Approaches Fall Short} \label{subsec:limitation}

In the following, we briefly review existing approaches on watermarking LLM-generated code.

Among existing LLM watermarking approaches, most efforts focus on AI-generated text detection~\citep{watermark-for-llm,giboulot2024watermax,zhaoprovable,liu2024adaptive,christ2024undetectable,zhang2024remark,dathathri2024scalable}.
As the state-of-the-art approaches, \citet{watermark-for-llm} firstly propose WLLM (Watermarking LLMs), which works by splitting a vocabulary into ‘green’ and ‘red’ tokens and softly promoting ‘green’ ones during text generation.
As the split of tokens is unknown to adversaries, WLLM then detects whether given text is AI-generated by identifying the presence of the ‘green’ tokens.
\citet{giboulot2024watermax} propose WaterMax, which embeds watermarks by generating multiple candidate text chunks from an LLM and selecting the one with particular distributions that maximize a predefined test statistic.
The watermark is then identified by evaluating whether the $p$-value of the statistic on given text is under a controlled false alarms.

As an emerging task, only limited watermarking approaches focus on AI-generated code detection.
\citet{who-wrote} propose a selective watermarking technique via entropy thresholding called SWEET, which extends WLLM by prompting ‘green’ tokens at high-entropy positions to preserve the utility of code.
\citet{kim2025marking} further extend SWEET and propose a syntax token preserving watermarking approach called STONE.
For watermarking, STONE excludes the tokens critical to language construction, so as to reduce functional degradations.

However, existing state-of-the-art approaches are far from practical, suffering from the following limitations.

\emph{Discriminability limitation}.
Based on our empirical evaluations (as in Section~\ref{subsec:discriminability}), these approaches demonstrate limited accuracy in AI-generated code detection.
In particular, WLLM, SWEET and STONE show limited ability to search the ‘green’ tokens for watermarking, as code vocabularies are scarcer and the fixed tokens provide limited watermarking space.
Similarly, WaterMax fails to produce particularly distributed watermarks in common code in contrast to text.

\emph{Efficiency limitation}.
Based on our empirical evaluations (as in Section~\ref{subsec:efficiency}), these approaches show significant time and memory consumption.
In particular, the LLM modification-based watermarking approaches (WLLM, SWEET and STONE) interfere with the optimization strategies of LLMs, resulting in additional overheads that slow down code generation.
WaterMax forces LLMs to perform extensive repeated generations, leading to time and memory consumption.

\emph{Universality limitation}.
Existing white-box watermarking approaches require LLM-specific updates or optimizations that access LLM internals when adapting to different LLMs, which makes them inapplicable to closed-source LLMs and shows limited universality.

Therefore, in this work, we aim to devise effective watermarking strategies that address existing limitations as well as satisfy our requirements.

\section{Our Approach}
\label{sec:approach}

\begin{figure*}[!t]
	\hfill
	\begin{center}
		\includegraphics[width=\textwidth]{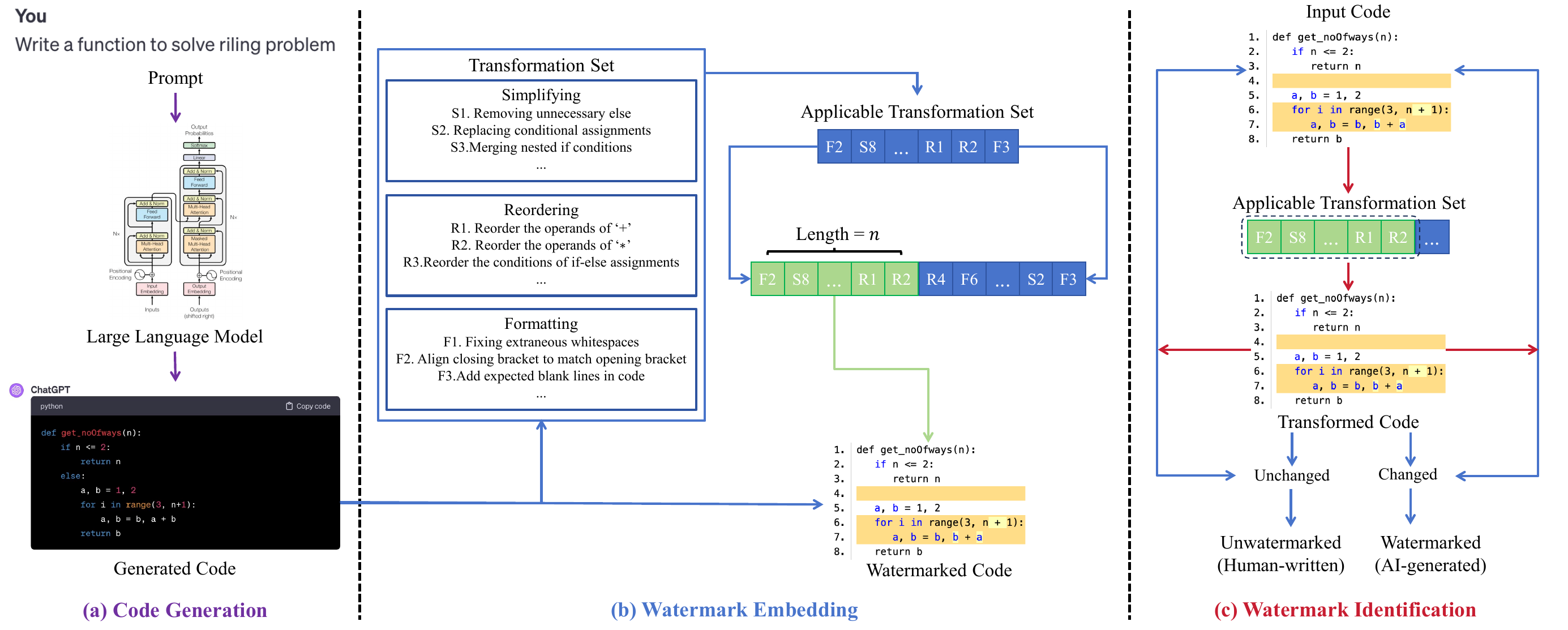}
	\end{center}
	\caption{An overview of \textbf{ACW}.}
    %\vspace{-5pt}
	\label{fig:framework}
\end{figure*}

In this section, we present \textbf{ACW} in detail, as overviewed in Figure~\ref{fig:framework}. 
The key idea of \textbf{ACW} is to selectively apply a set of carefully-designed code transformations for watermark embedding, and identify watermarks by checking the presence of our transformations.
If the watermark is identified to exist in a piece of code, it will be determined as AI-generated.

In the following, we first explain how the embedding function and the identification function work, and then present how the transformations are designed.

\subsection{Watermark Embedding} \label{subsec:embedding}

Given an AI-generated code snippet $\mathcal{C}$, and a set of code transformations $\mathcal{T}$, the embedding function $E$ works as follows.

\begin{algorithm}[t]
	\caption{Watermark Embedding}
	\small
    \begin{spacing}{1.2}
	\begin{algorithmic}[1]
		\STATE \textit{Input:} $C$ - code snippet
		\STATE \hspace{2.5em} $\mathcal{T}$ - a set of transformations
		\STATE \hspace{2.5em} $\omega$ - the watermark
		\STATE \hspace{2.5em} $n$ - hyperparameters
		\STATE \textit{Output:} $C^\omega$ - watermarked code snippet
		\STATE $T \gets \emptyset$
		\STATE $C^\omega \gets C$
		\FOR{$t \in \mathcal{T}$}
			\IF{$t$ is applicable to $C$}
				\STATE $T \text{.add}(t)$
			\ENDIF
		\ENDFOR
		\STATE $T \gets \text{SORT}(T)$
		\FOR{$t \in T[:n]$}
			\STATE $C^\omega \gets t(C^\omega)$
		\ENDFOR
		\RETURN $C^\omega$
	\end{algorithmic}
    \end{spacing}
	\label{alg:embedding}
\end{algorithm}

Firstly, the embedding function systematically analyzes $\mathcal{C}$ and identifies the code transformations in $\mathcal{T}$ that are applicable to $\mathcal{C}$. 
In order to efficiently determine whether a code transformation is applicable or not, the transformations in $\mathcal{T}$ must be designed such that their applicability can be determined statically (based on the abstract syntax trees (AST)~\citep{ast} in this work), rather than based on certain runtime information.
For example, one of our transformations in $\mathcal{T}$ is applicable, as long as there is an \texttt{if-then-else} statement in $\mathcal{C}$, i.e., its applicability is determined based on the language construct, rather than determining whether a certain branch of the \texttt{if-then-else} statement is infeasible or not.

Afterwards, let the set of applicable transformations be $T$, the embedding function sorts them based on a fixed order (e.g., the alphabetical order of the hashcode of transformations in this work). 
We reasonably assume that this order is unknown to adversaries and the ordering function is secret.
The ordering simplifies the design of code transformations, since different transformations may affect each other; applying the same set of transformations in different orders may result in different code, and adopting a fixed order for applicable transformations potentially ensures fixed results.

Lastly, the embedding function selectively applies $n$ applicable transformations in $T$ to AI-generated code, and checking whether these transformations have been applied or not allows us to tell whether the code is watermarked.

Algorithm~\ref{alg:embedding} details our watermarking embedding function $E$, with inputs including a code snippet $C$, a set of transformations $\mathcal{T}$, a watermark $\omega$ and a hyperparameter $n$. 
It initializes an empty set of applicable transformations $T$ and a variable $C^\omega$ to store the watermarked code at lines 6 and 7. 
From lines 8 to 12, it iterates through all possible transformations, adding those applicable ones (i.e., applicable to $C$) to the set $T$. 
Next, it sorts $T$ at line 13 and applies the selected $n$ transformations to $C^\omega$ from lines 14 to 16. 
Finally, it returns the watermarked code $C^\omega$ at line 17.

\subsection{Watermark Identification} \label{subsec:identification}

Given a code snippet $\mathcal{C}$ which may or may not be watermarked, and a set of code transformations $\mathcal{T}$, the identification function $I$ aims to identify whether $\mathcal{C}$ is watermarked, or report there is no watermark (i.e., the code is not the product of the embedding function in \textbf{ACW}).

Firstly, the identification function $I$ analyzes $\mathcal{C}$ to identify those transformations in $\mathcal{T}$ that are applicable to $\mathcal{C}$, denoted as $T$. 
We assume that the set of applicable transformations $T$ identified by $I$ is the same as those identified by the embedding function $E$. 
This is only the case if each transformation in $T$ is designed to be applicable if and only if a certain language construct is present in $\mathcal{C}$ and the language construct is not removed after the transformation.

Afterwards, based on the selected $n$ applicable transformations, the identification function $I$ determines whether there are watermarks by checking whether the transformations have been applied on $\mathcal{C}$ already.
In this work, the adopted transformations must be idempotent, i.e., $t(C) = t(t(C))$ for any code $C$ and $t$ in $\mathcal{T}$.
It means that applying the transformation once or multiple times results in the same code.
With this assumption, the identification function simply applies each of the $n$ applicable transformations and checks whether the code is modified or not after the transformation. 
If the code remains unchanged, it is determined that the transformation has been applied previously. 
Otherwise, it is determined not to be applied.

We remark a judgment on whether a certain transformation has been applied may be wrong, i.e., $\mathcal{C}$ was naturally written in a way as if the code were the result of the transformation. 
For example, in the case of a transformation that reorders two operands in an addition operation according to certain orders, the two operands may happen to have the `right' order, as if the transformation has been applied. Such a misjudgment is however unlikely if we set $n$ to be sufficiently large.
Let $t$ be an arbitrary transformation in $T$, and $Pr_{t,C}$ be the probability of a code snippet $C$ naturally satisfying $t(C) = C$. 
The probability of all $n$ transformations in $T$ denoted as $t_1, t_2, \cdots, t_n$,  naturally appear to be `applied' is thus $Pr_{t_1,C} * Pr_{t_2,C} * \cdots * Pr_{t_n,C}$, which is quite small if $n$ is sufficiently large. 
For instance, assuming that $Pr_{t,C}$ is 50\% for any transformation $t$, $n = 6$ would make the probability less than 1.6\%.
Thus, \textbf{ACW} is confident enough to identify the existence of watermarks.

\begin{algorithm}[!t]
	\caption{Watermark Identification}
	\small
    \begin{spacing}{1.2}
	\begin{algorithmic}[1]
		\STATE \textit{Input:} $C$ - code snippet
		\STATE \hspace{2.5em} $\mathcal{T}$ - a set of transformations
		\STATE \hspace{2.5em} $n$ - hyperparameters
		\STATE \textit{Output:} \texttt{Watermarked} or \texttt{Unwatermarked}
		\STATE $T \gets \emptyset$
		\FOR{$t \in \mathcal{T}$}
			\IF{$t$ is applicable to $C$}
				\STATE $T \text{.add}(t)$
			\ENDIF
		\ENDFOR
		\STATE $T \gets \text{SORT}(T)$
		\FOR{$t \in T[:n]$}
			\IF{$t(C) \neq C$}
				\RETURN \texttt{Unwatermarked}
			\ELSE
				\STATE $C \gets t(C)$
			\ENDIF
		\ENDFOR
		\RETURN \texttt{Watermarked}
	\end{algorithmic}
    \end{spacing}
	\label{alg:identification}
\end{algorithm}

Algorithm~\ref{alg:identification} details our watermark identification function $I$, with inputs including a code snippet $C$, a set of pre-defined transformations $\mathcal{T}$ and the hyperparameter $n$ in the watermark embedding function. 
The output is \texttt{Unwatermarked} if no watermark is detected, otherwise, the output is \texttt{Watermarked}. 
From lines 5 to 11, corresponding to the watermark embedding algorithm, it starts with initializing an empty set of applicable transformations $T$ and then traverses $\mathcal{T}$ to identify and sort the applicable transformations. 
From lines 12 to 18, it iterates through the selected $n$ applicable transformations to check the existence of watermarks. 
If the code is different before and after transformation $t$, it returns \texttt{Unwatermarked}.
Otherwise, it updates the code snippet with transformation $t$. 
If the code is identical to the input after applying all transformations, it returns \texttt{Watermarked}.

\subsection{Transformation} \label{subsec:transformation}

\begin{table*}[t]
\caption{Examples of Refactoring (R), Reordering (D) and Formatting (F) Transformations}
\begin{center}
\Huge
\renewcommand\arraystretch{1.25}
\resizebox{\linewidth}{!}
{
\begin{tabular}{cm{15cm}ll}
\toprule
Number
& \multicolumn{1}{c}{Rule Description}
& \multicolumn{1}{c}{Original Code Example}
& \multicolumn{1}{c}{Transformed Code Example}
\\ \midrule
R1
&
Removing unnecessary \texttt{else}.
&
\begin{tabular}{l}
\lstinline{def f(a=None):}
\\[-0.5ex]
\hspace{2em} \lstinline{if a is None:}
\\[-0.5ex]
\hspace{4em} \lstinline{return 1}
\\[-0.5ex]
\hspace{2em} \lstinline{else:}
\\[-0.5ex]
\hspace{4em} \lstinline{# some code here}
\end{tabular} 
&
\begin{tabular}{l}
\lstinline{def f(a=None):}
\\[-0.5ex]
\hspace{2em} \lstinline{if a is None:}
\\[-0.5ex]
\hspace{4em} \lstinline{return 1}
\\[-0.5ex]
\hspace{2em} \lstinline{# some code here}
\end{tabular}
\\ \midrule
R2
&
Converting \texttt{for} loops into list comprehensions.
&
\begin{tabular}{l}
\lstinline{cubes = []}
\\[-0.5ex]
\lstinline{for i in range(20):}
\\[-0.5ex]
\hspace{2em} \lstinline{cubes.append(i**3)}
\end{tabular}
&
\begin{tabular}{l}
\lstinline{cubes = [i**3 for i in range(20)]}
\end{tabular}
\\ \midrule
R3
&
Removing unnecessary casts to int, str, float and bool.
&
\begin{tabular}{l}
\lstinline{num_hats = 1} \\
\lstinline{update_hat_count(int(num_hats))}
\end{tabular}
&
\begin{tabular}{l}
\lstinline{num_hats = 1} \\
\lstinline{update_hat_count(num_hats)}
\end{tabular}
\\ \midrule
R4
&
Merging multiple comparisons into a single one.
&
\begin{tabular}{l}
\lstinline{if pay == "USD" or pay == "EUR":} \\
\hspace{2em} \lstinline{process(pay)}
\end{tabular}
&
\begin{tabular}{l}
\lstinline{if pay in ["USD", "EUR"]:} \\
\hspace{2em} \lstinline{process(pay)}
\end{tabular}
\\ \midrule
R5
&
Inlining a variable to a return in the case when the variable being declared is immediately returned.
&
\begin{tabular}{l}
\lstinline{def add_and_return(x, y):} \\
\hspace{2em} \lstinline{result = x + y} \\
\hspace{2em} \lstinline{return result}
\end{tabular}
&
\begin{tabular}{l}
\lstinline{def add_and_return(x, y):} \\
\hspace{2em} \lstinline{return x + y}
\end{tabular}
\\ \midrule
R6
&
Removing unnecessary slice indices.
&
\begin{tabular}{l}
\lstinline{numbers[0 : len(numbers)]}
\end{tabular}
&
\begin{tabular}{l}
\lstinline{numbers[:]}
\end{tabular}
\\ \midrule
R7
&
Computing the hash value of a certain class that contains more than one parameter.
If the hash value is an odd number, augmenting an unexecuted constructor, otherwise, remaining unchanged.
&
\begin{tabular}{l}
\lstinline{class A:}
\\
\hspace{2em} \lstinline{def __init__(self, x, y, z):}
\\[-0.5ex]
\lstinline{# Given Hash ("A") mod 2 = 1}
\end{tabular}
&
\begin{tabular}{l}
\lstinline{class A:}
\\
\hspace{2em} \lstinline{def __init__(self, x, y, z):}
\\
\hspace{2em} \lstinline{def __init__(slef, x): # New Constructor} 
\end{tabular}   
\\ \bottomrule
D1
&
Comparing the hash values of two operands involved in an ‘$+$’ operation.
Reordering them or remaining unchanged, ensuring operands with larger values always come before the smaller ones.
&
\begin{tabular}{l}
\lstinline{for i in range(3, n + 1):}
\\[-0.5ex]
\hspace{2em} \lstinline{a, b = b, a + b}
\\[-0.5ex]
\lstinline{return b}
\\[-0.5ex]
\lstinline{# Given Hash ("a") < Hash ("b")}
\end{tabular}
&
\begin{tabular}{l}
\lstinline{for i in range(3, n + 1):}
\\[-0.5ex]
\hspace{2em} \lstinline{a, b = b, b + a}
\\[-0.5ex]
\lstinline{return b}
\end{tabular}
\\ \midrule
D2
&
Comparing the hash values of two operands involved in an ‘$*$’ operation.
Reordering them or remaining unchanged, ensuring operands with larger values always come before the smaller ones.
&
\begin{tabular}{l}
\lstinline{for i in range(3, n + 1):}
\\[-0.5ex]
\hspace{2em} \lstinline{a, b = b, a * b}
\\[-0.5ex]
\lstinline{return b}
\\[-0.5ex]
\lstinline{# Given Hash ("a") < Hash ("b")}
\end{tabular}
&
\begin{tabular}{l}
\lstinline{for i in range(3, n + 1):}
\\[-0.5ex]
\hspace{2em} \lstinline{a, b = b, b * a}
\\[-0.5ex]
\lstinline{return b}
\end{tabular}
\\ \midrule
D3
&
Comparing the hash values of two operands involved in a ‘$>=$’ or ‘$<=$’ operator.
Reordering them or remaining unchanged, ensuring operands with larger values always come before the smaller ones.
&
\begin{tabular}{l}
\lstinline{if n <= left and n <= right:}
\\[-0.5ex]
\hspace{2em} \lstinline{return True}
\\[-0.5ex]
\lstinline{# Given Hash ("n") > Hash ("left")}
\\[-0.5ex]
\lstinline{# Given Hash ("n") < Hash ("right")}
\end{tabular}
&
\begin{tabular}{l}
\lstinline{if n <= left and right >= n:}
\\[-0.5ex]
\hspace{2em} \lstinline{return True}
\end{tabular}        
\\ \bottomrule
F1
&
Adding excepted blank lines.
&
\begin{tabular}{l}
\lstinline{def example_function1(param1, param2):}
\\[-0.5ex]
\hspace{2em} \lstinline{return param1 + param2}
\\[-0.5ex]
\lstinline{def example_function2(param1, param2):}
\\[-0.5ex]
\hspace{2em} \lstinline{return param1 - param2}
\end{tabular}
&
\begin{tabular}{l}
\lstinline{def example_function1(param1, param2):}
\\[-0.5ex]
\hspace{2em} \lstinline{return param1 + param2}
\\[-0.5ex]
\\[-0.5ex]
\lstinline{def example_function2(param1, param2):}
\\[-0.5ex]
\hspace{2em} \lstinline{return param1 - param2}
\end{tabular}
\\ \midrule
F2
& 
Fixing missing whitespaces around operators or keywords.
& 
\begin{tabular}{l}
\lstinline{def calculate(a, b, c, d):}
\\[-0.5ex]
\hspace{2em} \lstinline{result=a+b-c*d}
\\[-0.5ex]
\hspace{2em} \lstinline{return result}
\end{tabular}
&
\begin{tabular}{l}
\lstinline{def calculate(a, b, c, d):}
\\[-0.5ex]
\hspace{2em} \lstinline{result = a + b - c * d}
\\[-0.5ex]
\hspace{2em} \lstinline{return result}
\end{tabular}
\\ \midrule
F3
&
Aligning closing brackets to match opening brackets.
&
\begin{tabular}{l}
\lstinline{def example_function(param1, param2,}
\\[-0.5ex]
\hspace{2em} \lstinline{param3, param4):}
\\[-0.5ex]
\hspace{2em} \lstinline{return param1 + param2 + param3 + param4}
\end{tabular}
&
\begin{tabular}{l}
\lstinline{def example_function(param1, param2,}
\\[-0.5ex]
\hspace{10.25em} \lstinline{param3, param4):}
\\[-0.5ex]
\hspace{2em} \lstinline{return param1 + param2 + param3 + param4}
\end{tabular}
\\ \bottomrule
\end{tabular}
}
\label{tab:transformation}
\end{center}
\end{table*}

Our designed transformation set is a key component in \textbf{ACW}.  
In the following, we present how we design the transformations.

\textbf{ACW} currently supports 46 transformations categorized into three groups, including refactoring, reordering, and formatting rules.
Without loss of generality, we focus on watermarking AI-generated Python code in this work, and it is straightforward that \textbf{ACW} can be applied to other programming languages, as our semantic-preserving transformation set can be naturally expanded based on rich language features provided by modern programming languages.

Table~\ref{tab:transformation} presents an example transformation set in \textbf{ACW}, and the remaining transformations are presented in an online appendix (as in Section~\ref{sec:availability}).
We remark that many of the transformations are inspired by existing compilers or code optimizers~\citep{sourcery,autopep8}, and some of them are designed with certain `random' factors, such that it is challenging for an adversary to infer what transformations are supported and how they work.

\emph{Refactoring transformations} are motivated based on the fact that different programming syntax rules can often express the same semantics.
For example, a \texttt{for} loop can be transformed into a list expression that retains the same semantics. 
In addition to intuitive rules, this category also includes sophisticated transformations that leverage language features, such as function overloading.
For example, for a class with a constructor that takes multiple parameters, we overload the class constructor and introduce an additional one, which only occurs if the hash value of the class name is odd.

\emph{Reordering transformations} are motivated based on the fact that many operators exhibit commutativity and associativity, i.e., their operands can be reordered without altering code semantics. 
For example, a comparison expression \texttt{n $<=$ right} remains unchanged if the hash value of \texttt{n} is greater than or equal to that of \texttt{right}, otherwise, it will be transformed into \texttt{right $>=$ n}.
It is straightforward to see such a transformation can also be extended to other binary operators, such as arithmetic operators (e.g., addition, multiplication), or logical operators like conjunction and disjunction.

\emph{Formatting transformations} are motivated based on the fact that much human-written code may follow certain standards or styles, e.g., the PEP8 standard~\citep{pep8-standard}, and such phenomena can also exist in AI-generated code. 
Through unifying code standards and styles as well as rectifying potential formatting flaws, the produced code adheres to consistent `standardized' coding patterns that constitute discriminating watermarks.

%\subsection{Summary} \label{subsec:multibit}

In summary, our transformations are designed with multiple characteristics.
First, it should efficiently and statically determine whether a transformation is applicable or not. 
This is true since we can check whether a transformation is applicable by checking whether certain syntactic patterns exist in code. 
Second, the applicability of each transformation should remain unchanged after applying the transformation.
This is achieved by preserving the syntactic pattern after transformations and can be evident by examining each transformation. 
Third, all transformations are designed to be semantic-preserving and idempotent, which can be shown by examining the design of each transformation one by one. 
Lastly, it should be apparent that extending our transformation library with additional rules is straightforward.

\begin{figure}[t]
    \centering
    \includegraphics[width=0.75\linewidth]{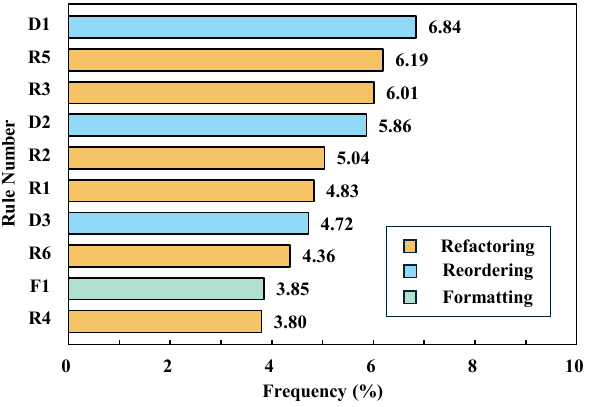}
    \vspace{-5pt}
    \caption{Statistics of the top-10 contributing transformation rules.}
    \vspace{-5pt}
    \label{fig:frequency}
\end{figure} 

Figure~\ref{fig:frequency} presents statistics of the top-10 applicable transformation rules contributing to code watermarking, where the rule numbering follows that in Table~\ref{tab:transformation}.
As in the figure, the columns in different colors (corresponding to different types of transformations) indicate the relative application frequency of these transformation rules, i.e., the proportion of the frequency each transformation is applied against the sum application frequency across all transformations (110606 times in total), which are counted based on the total of our experimental code in Section~\ref{subsec:setup}.
The top-10 relative application frequency ranges from 3.80\% to 6.84\%, which reflects the intended balanced application of different transformations in \textbf{ACW}.
We will further ablatively explore the effect of our transformation rules from multiple perspectives in Section~\ref{sec:study}.

\section{Experiment}
\label{sec:experiment}

In this section, we conduct a series of experiments to evaluate the effectiveness of \textbf{ACW}, by answering the following research questions (RQs).
\begin{itemize}
	\item
	\emph{RQ1 (Discriminability): Is \textbf{ACW} effective in distinguishing AI-generated code from human-written code?}
	\item
	\emph{RQ2 (Efficiency): Is \textbf{ACW} efficient in watermarking AI-generated code?}
    \item
	\emph{RQ3 (Utility): Does \textbf{ACW} preserve code utility?}
	\item
    \emph{RQ4 (Resilience): How resilient is \textbf{ACW} against potential code disruptions?}
\end{itemize}

We release our code and data to implement \textbf{ACW} available online, as in Section~\ref{sec:availability}.
All the experiments in this work are conducted on a server with an Intel\textsuperscript{\textregistered} Xeon\textsuperscript{\textregistered} Gold 6348 CPU (2.60GHz), 240GB of RAM, and an NVIDIA A800 GPU, 80GB of memory.

\subsection{Experiment Setup}
\label{subsec:setup}

\textbf{ACW} is implemented as a plug-and-play tool universal for different LLMs.
It currently includes a set of 46 transformations for Python code, and some of which are adopted from code optimization libraries such as \emph{Sourcery}~\citep{sourcery} and \emph{Autopep8}~\citep{autopep8}.
Those transformations containing randomness are implemented using a stable SHA-256~\citep{sha-256} hash algorithm (that is assumed to be replaceable and unknown to adversaries).

\emph{Datasets.} 
We collect an extensive set of data encompassing function-level and project-level Python code, which is generated using open-source and closed-source LLMs in combination with widely adopted prompt benchmarks, as well as multiple-source human-written code, enabling comprehensive evaluations.

We firstly generate a collection of function-level Python code based on the state-of-the-art LLMs including GPT-4~\citep{openai2023chatgpt}, GPT-4o and Qwen2.5-Coder-14B-FP16~\citep{hui2024qwen2}, with prompt datasets including MBPP~\citep{mbpp}, APPS~\citep{apps} and HumanEval+~\citep{liu2023codegeneratedchatgptreally}, which are widely adopted in code-related research~\citep{who-wrote,kim2025marking}.
Using each LLM, we generate 974, 3035, and 164 pieces of code based on MBPP, APPS and HumanEval+, respectively, producing nine sets of AI-generated code, including MBPP-GPT-4, MBPP-GPT-4o, MBPP-Qwen, APPS-GPT-4, APPS-GPT-4o, APPS-Qwen, HumanEval-GPT-4, HumanEval-GPT-4o and HumanEval-Qwen.
We systematically apply \textbf{ACW} to watermark these AI-generated code, and assign each set with the same quantity of human-written code, which is collected from the corresponding benchmark data provided by MBPP, APPS and HumanEval+. 

Additionally, we generate a collection of Python projects based on GPT-4, GPT-4o, and a representative LLM agent, ChatDev~\citep{qian2024chatdev}.
ChatDev incorporates a Software Requirement Description Dataset (SRDD), which is the largest project-level prompt dataset to date~\cite{LiuPHHELDG25}.
Qwen2.5-Coder-14B-FP16 is not adopted as the context processing capability of an open-source LLM cannot support large-scale information flows and multi-round decision-making required for agents.
Based on each LLM, we generate 500 projects that functionally decompose 7465 executable-function code, and produce SRDD-GPT-4 and SRDD-GPT-4o testing sets.
We systematically apply \textbf{ACW} to watermark these AI-generated projects through watermarking each executable function, and assign each set with the same quantity of human-written code projects from a representative CodeSearchNet~\cite{husain2019codesearchnet} challenge dataset.

\emph{Baseline Methods.}
We apply the state-of-the-art AI-generated content detection methods based on watermarking, including WLLM~\citep{watermark-for-llm}, WaterMax~\citep{giboulot2024watermax}, SWEET~\citep{who-wrote} and STONE~\citep{kim2025marking} as our baselines, as detailed in Section~\ref{subsec:limitation}.
We re-implement the baselines based on the source code provided in~\citep{kim2025marking} (WLLM, SWEET and STONE) and~\citep{giboulot2024watermax} (WaterMax), with hyperparameters identical to the original settings in the literature.
Note that we do not conduct any re-training or fine-tuning for fair comparisons.
In addition, we evaluate these baselines based on the Qwen LLM, as their white-box watermarking strategies do not apply to those closed-source LLMs such as ChatGPT.

\emph{Metrics.}
We measure the performance of \textbf{ACW} and other baselines in distinguishing AI-generated (a.k.a. positive) or human-written (a.k.a. negative) code, based on the following evaluation metrics.

\begin{itemize}
    \item $ACC = (TP + TN) / (TP + TN + FP + FN)$ measures the overall performance, where $TP + TN$ refers to the total correctly-identified code and $TP + TN + FP + FN$ refers to the total code.
    \item $TPR = Recall = TP / (TP + FN)$ measures the performance in identifying watermarks, where $TP$ refers to the correctly-identified AI-generated code and $TP + FN$ refers to the total AI-generated code. 
    \item $FPR = FP / (FP + TN)$ measures false alarms, where $FP$ refers to the incorrectly-identified human-written code and $FP + TN$ refers to the total human-written code.
    \item $Precision = TP / (TP + FP)$ measures the reliability in avoiding false alarms, where $TP + FP$ refers to the total code that is identified as AI-generated.
    \item $F1\text{-Score} = 2 \times (Precision \times Recall) /(Precision + Recall)$ measures the balanced performance on identifying watermarks and avoiding false alarms.
\end{itemize}
Note that we determine an AI-generated code project through binomial tests on the watermark identification of its executable functions, with a strict significance threshold at 0.05.

We measure the efficiency of \textbf{ACW} and other baselines in terms of the time (in seconds) and GPU memory (in mebibytes) cost for watermarking.
In addition, we measure the utility of AI-generated code before and after watermarking in terms of Pass Rate, i.e., the proportion of test-passing samples among the total.
The function-level pass rate is evaluated based on the first-generation code that successfully passes the test cases provided by the corresponding prompt datasets, MBPP, APPS and HumanEval+.
The project-level pass rate is evaluated based on the executable functions in code projects that are successfully parsed by syntax trees~\cite{treesitter} while ensuring all the syntax tree nodes are valid.

\begin{table*}[t]
    \caption{Evaluation Results on Discriminability}
	\begin{center}
        \large
		\renewcommand\arraystretch{1.05}
		{\resizebox{0.7\textwidth}{!}{
			\centering
                \begin{tabular}{ccccccc} 
                    \toprule
                    Method & Dataset & ACC (\%) & TPR (\%) & FPR (\%) 
                    & Precision (\%) & F1-Score (\%) \\ 
                    \midrule
                    \multirow{11}{*}{\textbf{ACW}}   
                    & MBPP-GPT-4 
                    & 98.97 & 97.94 & 0 & 100 & 98.96 \\
                    & APPS-GPT-4
                    & 98.43 & 97.03 & 0.17 & 99.83 & 98.41 \\
                    & HumanEval-GPT-4     
                    & 97.56 & 95.12 & 0 & 100 & 97.50 \\
                    & SRDD-GPT-4         
                    & 100 & 100 & 0 & 100 & 100 \\
                    & MBPP-GPT-4o         
                    & 98.51 & 97.02 & 0 & 100 & 98.49 \\
                    & APPS-GPT-4o         
                    & 98.14 & 96.44 & 0.17 & 99.82 & 98.10 \\
                    & HumanEval-GPT-4o    
                    & 97.87 & 95.73 & 0 & 100 & 97.82 \\
                    & SRDD-GPT-4o         
                    & 100 & 100 & 0 & 100 & 100 \\
                    & MBPP-Qwen      
                    & \textbf{98.81} & \textbf{97.63} & \textbf{0} & \textbf{100} & \textbf{98.80} \\
                    & APPS-Qwen      
                    & \textbf{98.15} & \textbf{96.47} & \textbf{0.17} & \textbf{99.82} & \textbf{98.12} \\
                    & HumanEval-Qwen 
                    & \textbf{97.25} & \textbf{94.51} & \textbf{0} & \textbf{100} & \textbf{97.18} \\ 
                    \midrule
                    \multirow{3}{*}{WLLM}  
                    & MBPP-Qwen      
                    & 68.08 & 62.98 & 26.82 & 70.13 & 66.36 \\
                    & APPS-Qwen      
                    & 68.20 & 67.32 & 30.92 & 68.53 & 67.92 \\
                    & HumanEval-Qwen 
                    & 68.26 & 70.22 & 33.70 & 67.57 & 68.87 \\ 
                    \midrule
                    \multirow{3}{*}{WaterMax}  
                    & MBPP-Qwen      
                    & 57.23 & 19.29 & 4.83 & 79.98 & 31.08 \\
                    & APPS-Qwen      
                    & 59.27 & 23.35 & 4.96 & 82.48 & 36.40 \\
                    & HumanEval-Qwen 
                    & 59.27 & 22.80 & 4.27 & 84.23 & 35.89 \\ 
                    \midrule
                    \multirow{3}{*}{SWEET} 
                    & MBPP-Qwen      
                    & 75.64 & 55.91 & 4.62 & 92.37 & 69.66 \\
                    & APPS-Qwen      
                    & 71.20 & 47.80 & 5.40 & 89.85 & 62.40 \\
                    & HumanEval-Qwen 
                    & 74.67 & 49.95 & 0.60 & 98.81 & 66.36 \\ 
                    \midrule
                    \multirow{3}{*}{STONE} 
                    & MBPP-Qwen      
                    & 70.84 & 64.88 & 23.20 & 73.66 & 68.99 \\
                    & APPS-Qwen      
                    & 68.36 & 69.98 & 33.26 & 67.78 & 68.86 \\
                    & HumanEval-Qwen 
                    & 76.52 & 66.46 & 13.41 & 83.21 & 73.90 \\
                    \bottomrule
                \end{tabular}
                }}
	\end{center}
    \label{tab:discriminability}
\end{table*}

\begin{figure}[!b]
    \centering
    \vspace{-10pt}
    \includegraphics[width=0.75\linewidth]{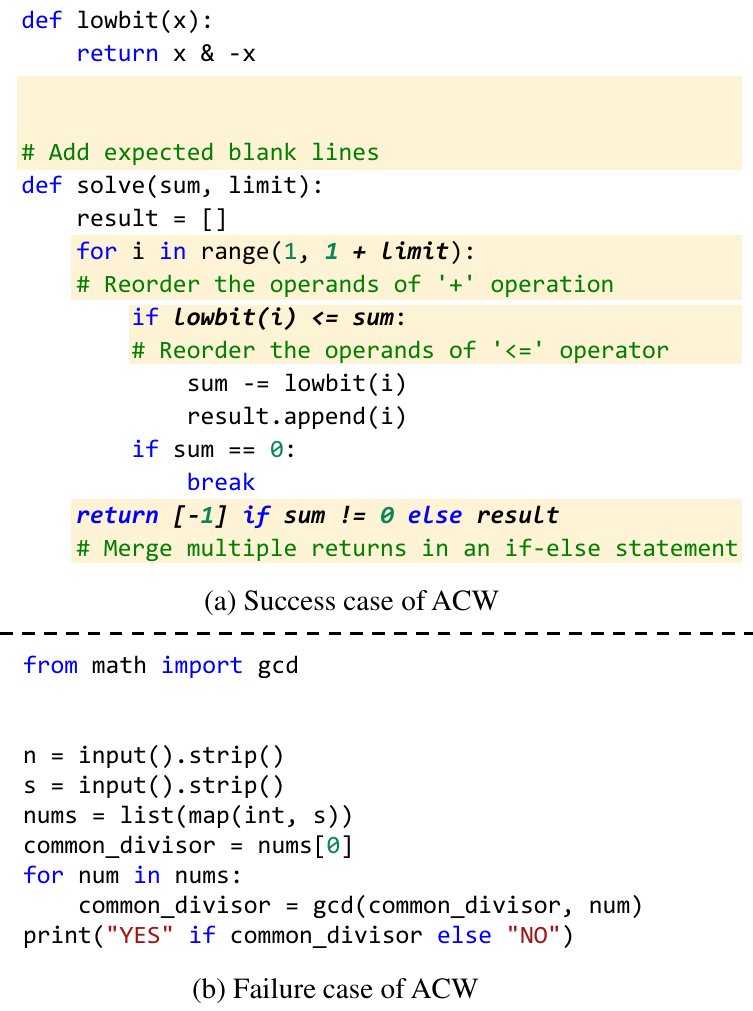}
    \caption{Example cases of ACW, where the transformed (watermarked) elements are highlighted with italics and yellow backgrounds.}
    \label{fig:caseacw}
\end{figure}

\subsection{RQ1: Discriminability}
\label{subsec:discriminability}

Through this experiment, we aim to answer: \emph{Is \textbf{ACW} effective in distinguishing AI-generated code from human-written code?}

In the following, we evaluate \textbf{ACW} and other baselines in distinguishing AI-generated code.
Table~\ref{tab:discriminability} presents our evaluation results.
Based on merged AI-generated and human-written code datasets, we systematically watermark the generated code using \textbf{ACW} and other baselines, and then identify the watermark to evaluate the discriminability in terms of ACC, TPR, FPR, Precision and F1-Score.
For \textbf{ACW}, we set $n = |T|$, i.e., a general setting in which all the applicable transformations are utilized for evaluation.
For the baseline methods, we present their optimal results when testing with different thresholds.

As in Table~\ref{tab:discriminability}, \textbf{ACW} demonstrates promising discriminability and outperforms other baselines (as the best results in bold), with ACC over 97\%, TPR over 94\%, FPR below 1\%, Precision over 99\% and F1-Score over 97\% on all datasets.
For example, based on MBPP-Qwen, \textbf{ACW} achieves ACC of 98.81\%, TPR of 97.63\%, FPR of 0\%, Precision of 100\% and F1-Score of 98.80\%.
The promising results demonstrate the effectiveness of \textbf{ACW} in distinguishing AI-generated code from human-written merged sets (in terms of ACC), with balanced performance (in terms of F1-Score) on identifying watermarks (in terms of TPR) and avoiding false alarms (in terms of FPR and Precision).
To qualitatively analyze the success and failure cases of \textbf{ACW}, as exemplified in Figure~\ref{fig:caseacw}, the limited failures are caused by some overly simple code that involves almost no logic for semantic-preserving transformations.
Note that the overly simple code poses limited authenticity concerns and is commonly considered less interesting for watermarking.
Such phenomena do not appear on the project-level SRDD-GPT-4 and SRDD-GPT-4o datasets (with outstanding TPR of 100\%), as complex code projects provide more space for watermarking compared to code functions.

\begin{figure*}[b]
    \centering
    \includegraphics[width=0.98\linewidth]{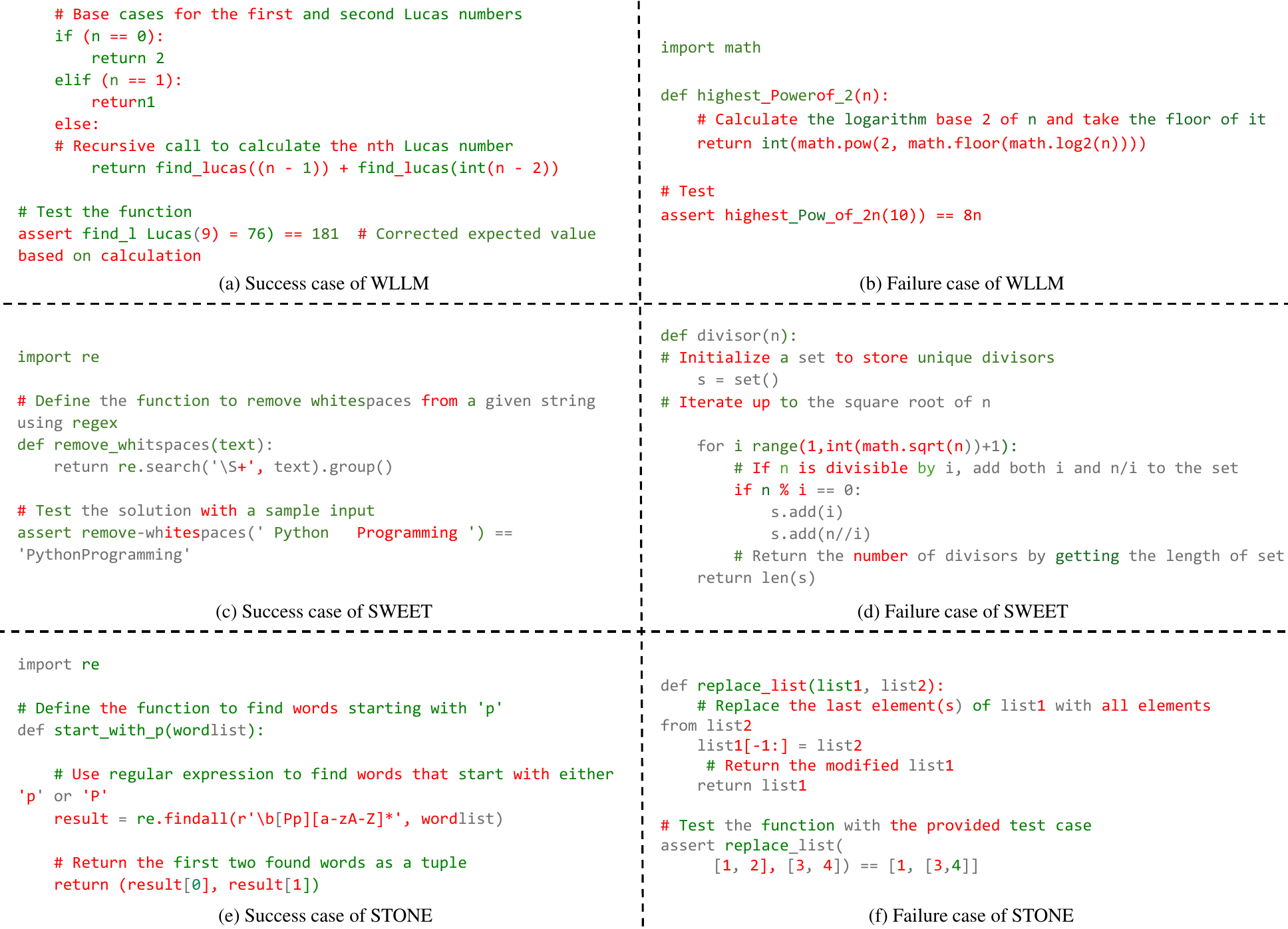}
    \vspace{-5pt}
    \caption{Example cases of WLLM, SWEET and STONE, where the ‘green’ tokens are the candidates modified for watermarking, while the ‘red’ tokens and the excluded ‘gray’ tokens are not involved in the watermarking process.}
    %\vspace{-10pt}
    \label{fig:casewllm}
\end{figure*}

Compared with \textbf{ACW}, these baselines present unsatisfactory performance.
WLLM~\citep{watermark-for-llm} and WaterMax~\cite{giboulot2024watermax} report promising results on detecting AI-generated text, but the same is not true for code.
For example, based on MBPP-Qwen, WLLM achieves ACC of 68.08\%, TPR of 62.98\%, FPR of 26.82\%, Precision of 70.13\%, F1-Score of 66.36\%, and WaterMax achieves ACC of 57.23\%, TPR of 19.29\%, FPR of 4.83\%, Precision of 79.98\%, F1-Score of 31.08\%.
SWEET~\citep{who-wrote} and STONE~\citep{kim2025marking} attempt to improve WLLM through adjusting the token selection strategy for adapting to code, but show limited success.
For example, based on MBPP-Qwen, SWEET achieves ACC of 75.64\%, TPR of 55.91\%, FPR of 4.62\%, and STONE achieves ACC of 70.84\%, TPR of 64.88\%, FPR of 23.20\%.

\begin{figure}[t]
    \centering
    \includegraphics[width=0.98\linewidth]{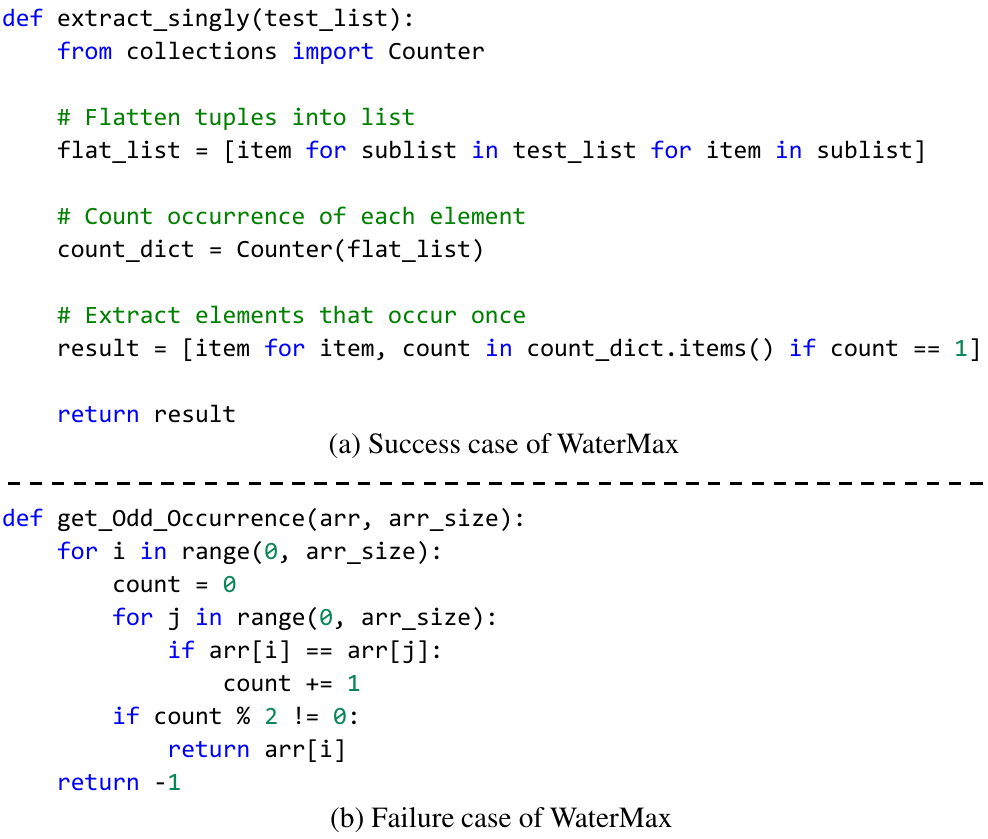}
    \caption{Example cases of WaterMax, where the watermarks are embedded in the AI-generated code with predefined statistical distributions.}
    \vspace{-5pt}
    \label{fig:casewatermax}
\end{figure}

We then qualitatively analyze the watermarking strategies of these baselines (as in Section~\ref{subsec:limitation}) based on their success and failure cases.
Note that we illustrate the example cases and provide the complete resultant data online (as in Section~\ref{sec:availability}).
As exemplified in Figure~\ref{fig:casewllm}, WLLM, SWEET and STONE show limited ability to search the ‘green’ tokens for watermarking, as code vocabularies are scarcer and the produced fixed tokens provide limited watermarking space.
They even tend to prompt ‘green’ tokens in code comments, which deviates from the goal of code watermarking and lacks practicality in real-world applications.
Similarly, as exemplified in Figure~\ref{fig:casewatermax}, WaterMax achieves limited success (with TPR below 25\% on all datasets) in the code with a high density of comments or close to text, but fails to produce particularly distributed watermarks in common code that differs from text.
In general, code presents lower flexibility than text and remains challenging for existing LLM watermarking approaches on AI-generated content detection.

\emph{Based on this experiment, we conclude that \textbf{ACW} is effective in distinguishing AI-generated code from human-written code.}

\subsection{RQ2: Efficiency}
\label{subsec:efficiency}

Through this experiment, we aim to answer: \emph{Is \textbf{ACW} efficient in watermarking AI-generated code?}

In the following, we evaluate \textbf{ACW} and other baselines in terms of time (in seconds) and GPU memory (in mebibytes) cost for watermarking, i.e., evaluating whether these watermarking strategies slow down LLM code generation.

Table~\ref{tab:efficiency} presents our evaluation results.
In the table, the results outside the brackets indicate the average time and GPU memory on watermarking one piece of code, and the results inside the brackets indicate the corresponding average cost on code generation.
Note that the code generation memory required by ChatGPT-related LLMs is not presented, as the closed-source LLMs are running online and the GPU memory is inaccessible.
In these baseline methods, the generating and watermarking processes are integrated.
For clear observations, we count the watermarking cost based on the difference between the integrated cost and the code generation cost (obtained based on the original LLMs without watermarking strategies applied).

As in Table~\ref{tab:efficiency}, \textbf{ACW} demonstrates promising efficiency and outperforms other baselines (as the best results in bold).
In terms of watermarking time, \textbf{ACW} costs less than 0.1 seconds on all datasets.
For example, based on MBPP-Qwen, \textbf{ACW} costs 0.03 seconds to watermark one piece of code, which is much less than the code generation time (1.10 seconds).
Based on SRDD-GPT-4, it costs an average of 0.83 seconds to watermark one code project, which is also negligible compared to generating a project (160.92 seconds).
In terms of GPU memory, \textbf{ACW} does not produce any additional overhead.
For example, based on MBPP-Qwen, the memory for generating one piece of code remains unchanged (29672 MiB) whether \textbf{ACW} is applied or not.
The promising results benefit from the post-processing mechanism of \textbf{ACW}, where static code transformations require minimal execution time and do not introduce additional GPU overhead for watermarking.

Compared with \textbf{ACW}, these baselines present unsatisfactory performance.
In terms of time, the watermarking cost of all baselines greatly exceeds the cost for code generation.
For example, based on MBPP-Qwen, SWEET costs 6.03 seconds for watermarking, which is much more than the time to generate one piece of code (1.10 seconds) and significantly slows down LLMs.
In terms of GPU memory, all the baselines introduce significant overheads.
Especially, SWEET requires excessive GPU memory to deal with an LLM with complicated prompts. 
For example, based on APPS-Qwen, SWEET requires 42160 MiB of memory to watermark one piece of code, which exceeds the memory cost for code generation (29783 MiB).
To further analyze the watermarking strategies of these baselines, WLLM, SWEET, and STONE introduce modifications to LLMs that disrupt LLM optimization strategies, leading to a modified LLM with lower efficiency than its original version.
WaterMax forces LLMs to perform exponentially repeated generations to produce outputs with particular distributions, resulting in significant time and memory consumption.
In contrast, \textbf{ACW} does not hinder LLM generations and maintains efficiency, providing practical watermarking solutions in real-world applications.

\begin{table}[t]
\caption{Evaluation Results on Efficiency}
\begin{center}
\Large
\renewcommand\arraystretch{1.1}
\resizebox{0.48\textwidth}{!}{
\begin{tabular}{cccc}
\toprule
Method                             
& Dataset        & Time Cost (s)   & GPU Memory (MiB)   \\ 
\midrule
\multirow{11}{*}{\textbf{ACW}} 
& MBPP-GPT-4       &  0.05 (3.08)   & 0 (N/A)  \\
& APPS-GPT-4       &  0.07 (3.28)   & 0 (N/A)  \\
& HumanEval-GPT-4  &  0.06 (2.92)   & 0 (N/A)  \\
& SRDD-GPT-4       &  0.83 (160.92) & 0 (N/A)  \\
& MBPP-GPT-4o      &  0.02 (2.96)   & 0 (N/A)  \\
& APPS-GPT-4o      &  0.05 (3.10)   & 0 (N/A)  \\
& HumanEval-GPT-4o &  0.06 (2.81)   & 0 (N/A)  \\
& SRDD-GPT-4o      &  0.79 (127.31) & 0 (N/A)  \\
& MBPP-Qwen      & \textbf{0.03 (1.10)} & \textbf{0 (29672)} \\
& APPS-Qwen      & \textbf{0.08 (2.01)} & \textbf{0 (29783)} \\
& HumanEval-Qwen & \textbf{0.09 (1.68)} & \textbf{0 (29217)} \\
\midrule
\multirow{3}{*}{WLLM}         
& MBPP-Qwen      & 6.83 (1.10)          & 27611 (29672)      \\
& APPS-Qwen      & 6.42 (2.01)          & 28374 (29783)      \\
& HumanEval-Qwen & 8.94 (1.68)          & 28032 (29217)      \\
\midrule
\multirow{3}{*}{WaterMax}         
& MBPP-Qwen      & 9.78 (1.10)          & 8592 (29672)      \\
& APPS-Qwen      & 10.42 (2.01)         & 10280 (29783)      \\
& HumanEval-Qwen & 9.69 (1.68)          & 8385 (29217)      \\
\midrule
\multirow{3}{*}{SWEET}        
& MBPP-Qwen      & 6.03 (1.10)          & 33917 (29672)      \\
& APPS-Qwen      & 5.10 (2.01)          & 42160 (29783)      \\
& HumanEval-Qwen & 7.79 (1.68)          & 32836 (29217)      \\
\midrule
\multirow{3}{*}{STONE}        
& MBPP-Qwen      & 6.54 (1.10)          & 27589 (29672)      \\
& APPS-Qwen      & 6.25 (2.01)          & 28288 (29783)      \\
& HumanEval-Qwen & 8.97 (1.68)          & 27964 (29217)      \\ 
\bottomrule
\end{tabular}}
\end{center}
\label{tab:efficiency}
\vspace{-5pt}
\end{table}

\emph{Based on this experiment, we conclude that \textbf{ACW} is efficient in watermarking AI-generated code.}

\subsection{RQ3: Utility}

\begin{table}[!t]
\caption{Evaluation Results on Utility}
\begin{center}
\Large
\renewcommand\arraystretch{1.1}
\resizebox{0.48\textwidth}{!}{
\begin{tabular}{ccccc}
\toprule
\multirow{2}{*}{Method}       
& \multirow{2}{*}{Dataset} 
& \multicolumn{2}{c}{Pass Rate (\%)} 
& \multirow{2}{*}{Degration (\%)} \\ \cmidrule(lr){3-4}
&                          
& Original   
& Watermarked   
&  \\ 
\midrule 
\multirow{11}{*}{\textbf{ACW}} 
& MBPP-GPT-4               
& 71.91           & 70.21              & 1.70                            
\\
& APPS-GPT-4               
& 54.79           & 54.33              & 0.46                            
\\
& HumanEval-GPT-4          
& 89.60           & 88.40              & 1.20                            
\\
& SRDD-GPT-4         
& 100 & 100 & 0
\\
& MBPP-GPT-4o              
& 82.13           & 81.49              & 0.64                            
\\
& APPS-GPT-4o              
& 59.93           & 59.64              & 0.29                            
\\
& HumanEval-GPT-4o         
& 83.50           & 82.80              & 0.70                            
\\
& SRDD-GPT-4o         
& 100 & 100 & 0
\\
& MBPP-Qwen                
& \textbf{63.12}  & \textbf{62.29}     & \textbf{0.83}                            
\\
& APPS-Qwen                
& \textbf{49.52}  & \textbf{48.40}     & \textbf{1.12}                            
\\
& HumanEval-Qwen           
& \textbf{76.80}  & \textbf{76.20}     & \textbf{0.60}                            
\\
\midrule 
\multirow{3}{*}{WLLM}    
& MBPP-Qwen                
& 63.12           & 28.64              & 34.48                           
\\
& APPS-Qwen                
& 49.52           & 17.31              & 32.21                           
\\
& HumanEval-Qwen           
& 76.80           & 27.43              & 49.37        
\\
\midrule 
\multirow{3}{*}{WaterMax}    
& MBPP-Qwen                
& 63.12           & 61.90              & 1.22                           
\\
& APPS-Qwen                
& 49.52           & 46.06              & 3.46                           
\\
& HumanEval-Qwen           
& 76.80           & 73.17              & 3.63        
\\
\midrule 
\multirow{3}{*}{SWEET}   
& MBPP-Qwen                
& 63.12           & 43.21              & 19.91                           
\\
& APPS-Qwen                
& 49.52           & 20.12              & 29.40                           
\\
& HumanEval-Qwen           
& 76.80           & 43.90              & 32.90                           
\\
\midrule 
\multirow{3}{*}{STONE}   
& MBPP-Qwen                
& 63.12           & 44.66              & 18.46                           
\\
& APPS-Qwen                
& 49.52           & 22.61              & 26.91                           
\\
& HumanEval-Qwen           
& 76.80           & 51.83              & 24.97              
\\ \bottomrule 
\end{tabular}}
\end{center}
\label{tab:utility}
\vspace{-5pt}
\end{table}

Through this experiment, we aim to answer: \emph{Does \textbf{ACW} preserve code utility?}

In the following, we evaluate the effect of \textbf{ACW} and other baselines on the utility of the AI-generated code before and after watermarking.
Table~\ref{tab:utility} presents our results, showing the pass rates of the original and watermarked code sets, and the degradation based on the pass rate difference between the two sets.
Note that the pass rates of the original code are not 100\% due to the limited code generation capability of LLMs.

As in Table~\ref{tab:utility}, \textbf{ACW} successfully preserves the utility of AI-generated code and outperforms other baselines (as the best results in bold).
Especially, the pass rate degradation of \textbf{ACW} is less than 2\% on all datasets.
For example, based on MBPP-Qwen, the pass rate of the watermarked code is 62.29\%, which is degraded by only 0.83\% compared to the original pass rate (63.12\%).
The promising results benefit from our design that all the transformations in \textbf{ACW} are semantic-preserving, and there will inevitably be minor errors resulting from unplanned transformation processes or runtime exceptions on test cases.

Compared with \textbf{ACW}, these baselines show clear impacts on the utility of AI-generated code.
For example, based on MBPP-Qwen, SWEET degrades the pass rate from 63.12\% to 43.21\%, with a degradation of 19.91\%.
To further analyze the watermarking process, WLLM, SWEET and STONE have promoted LLMs to generate text-driven code that involves syntax errors.
For example, in Figure~\ref{fig:casewllm} (c), the generated function names \texttt{remove\_whitspaces} and \texttt{remove-whitespaces} are inconsistent in the same code snippet.
WaterMax preserves code utility (with degradations below 4\%) as it almost fails to produce code watermarks (with TPR below 25\% on all datasets, as in Table~\ref{tab:discriminability}) and has no impact on code generation.
In general, the optimization of LLMs is sensitive and could be disrupted by LLM modifications, thereby degrading the code generation capability of LLMs.

We additionally evaluate the impact of \textbf{ACW} on large-scale human-built code projects.
For standard and fair evaluations, we utilize the project data, test environments and test cases provided by Tests4Py~\cite{smytzek2024tests4py}, an advanced benchmark for system testing.
Based on 507 projects (on average 127582 lines of code per project) corresponding to 507 commit versions collected from 21 Python projects, we systematically apply \textbf{ACW} to them and evaluate the utility before and after watermarking.
In particular, we conduct system and unit tests on each project using 100 test cases (i.e., 50700 test cases per test type), where the system test assesses the correctness of project initializations, and the unit test verifies whether the committed issues are resolved.
Consequently, 93.31\% of the system test cases and 99.20\% of the unit test cases are passed on the original projects, and the results remain unchanged after watermarking, further demonstrating the reliability of \textbf{ACW} on preserving code utility.

\emph{Based on this experiment, we conclude that \textbf{ACW} preserves the utility of AI-generated code successfully.}

\subsection{RQ4: Resilience}
\label{subsec:resilience}

Through this experiment, we aim to answer: \emph{How resilient is \textbf{ACW} against potential code disruptions?}

In the following, we evaluate the resilience of \textbf{ACW} by designing four potential attackers who aim to
(1) disrupt our watermark through post-optimizations on code, 
(2) disrupt our watermark through content-focused code modifications, 
(3) disrupt our watermark through adaptive attacks, and
(4) detect the existence of our watermark.

In particular, we implement the disruptive \emph{Attacker 1-3} by applying the corresponding disruptions on both AI-generated (watermarked) and human-written code, and comprehensively evaluate the resilience of \textbf{ACW} from three dimensions, including (1) TPR indicates the discriminability of our watermark with attacks, (2) FPR indicates false alarms that may result from human-written code forgeries, and (3) Pass Rate indicates attack validity, i.e., a valid attack is required to preserve code utility to be consistent with real-world threats.
We then implement the \emph{Attacker 4} by measuring the syntactic and functional similarity between the original and watermarked code, i.e., evaluating the imperceptibility of our watermark.

\begin{figure}[t]
    \centering
    \includegraphics[width=1\linewidth]{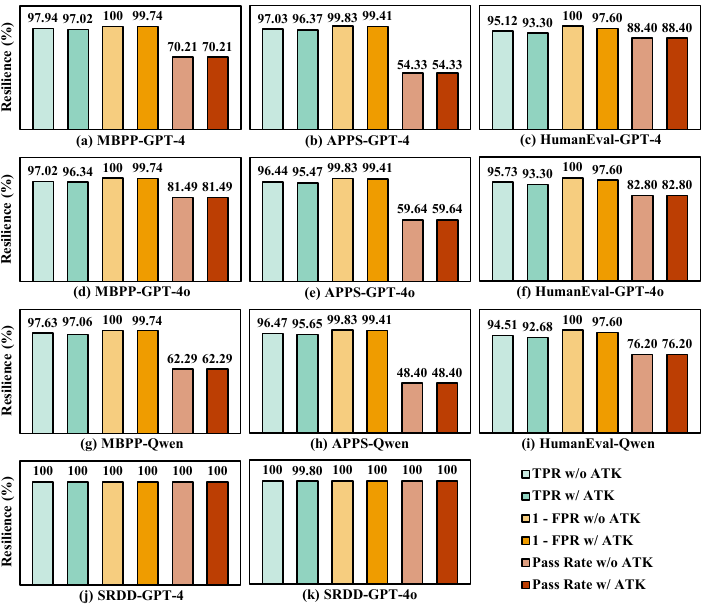}
    %\vspace{-10pt}
    \caption{Resilience results on default-level code optimizations.}
    \vspace{-5pt}
    \label{fig:attack1-1}
\end{figure}

\begin{figure}[t]
    \centering
    \includegraphics[width=1\linewidth]{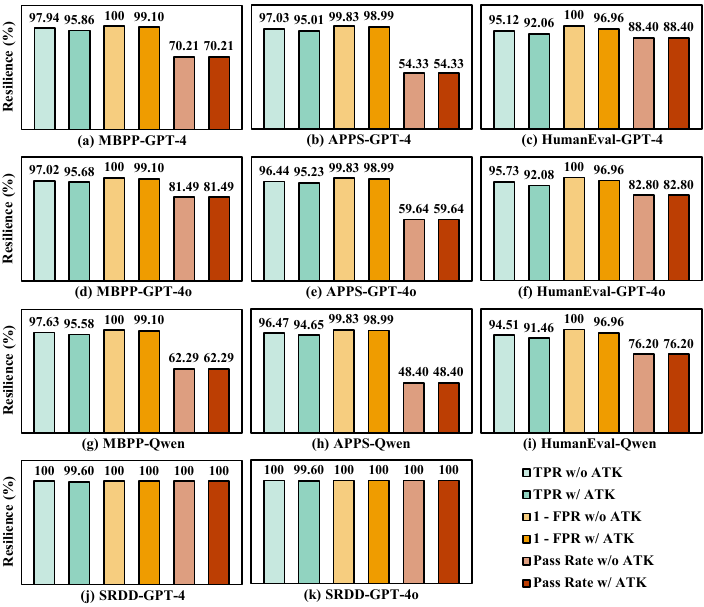}
    %\vspace{-10pt}
    \caption{Resilience results on maximum-level code optimizations.}
    \vspace{-5pt}
    \label{fig:attack1-2}
\end{figure}

\emph{Attacker~1: Code optimizations.}
The attacker aims to disrupt our watermark through post-optimizations on code, reflecting a real-world threat scenario where code is often optimized by compilers during programming.
We implement the attacker using an integrated code transformation toolkit, \emph{Ruff}~\cite{Ruff}, which integrates a wide range of libraries and transformation rules for Python code.
In particular, we adopt two settings in \emph{Ruff} for different levels of attacks, i.e., (1) a default level involving 103 transformations from \emph{Pyflakes}~\cite{pyflakes} and \emph{Pycodestyle}~\cite{pycode}, and (2) a maximum level involving 369 transformations from \emph{Pyflakes}, \emph{Pycodestyle}, \emph{Pylint}~\cite{pylint}, \emph{Pyupgrade}~\cite{pyupgrade}, \emph{Flake8-bugbear}~\cite{flake8-bugbear}, \emph{Flake8-simplify}~\cite{flake8-simplify}, \emph{Isort}~\cite{isort} and \emph{Refurb}~\cite{refurb}.

Figure~\ref{fig:attack1-1} and Figure~\ref{fig:attack1-2} present the default-level and maximum-level results on different datasets.
In each chart, the blue, orange and red columns respectively present the TPR, FPR and pass rate results with (w/) and without (w/o) attacks (ATK).
For clear observations, we present $1-$FPR results in the orange columns.
As in the charts, \textbf{ACW} achieves promising results under different levels of optimizations, maintaining both TPR and FPR on all datasets without pass rate degradations.
For example, in Figure~\ref{fig:attack1-1} (a), \textbf{ACW} achieves TPR of 97.02\% and $1-$FPR of 99.74\% on MBPP-GPT-4 under the default-level attack, showing negligible degradations compared to the unattacked TPR of 97.94\% and $1-$FPR of 100\%, and retaining the pass rate of 70.21\% unchanged.
The promising results demonstrate that our watermarks are discriminative enough to defend against code optimizations (in terms of TPR), as \textbf{ACW} contains abundant transformation rules to maintain the discriminability, even if some transformations are potentially forged to exist in human-written code (in terms of FPR).

\emph{Attacker~2: Content-focused code modifications.}
The attacker aims to disrupt our watermark through content-focused code modifications, following a forceful attack~\cite{pang2024no,pang2024attacking} that disrupts watermarks by modifying LLM-output words.
To reproduce the attacking strategy on code data, we modify specific content of code involving variable and function names using ChatGPT-4 with the prompt:
\emph{Rename the local variable and function names in the following Python code.
You must strictly preserve the input argument names and return values, and do not alter the logic, structure, or implementation details of the code.}

\begin{figure}[t]
    \centering
    \includegraphics[width=1\linewidth]{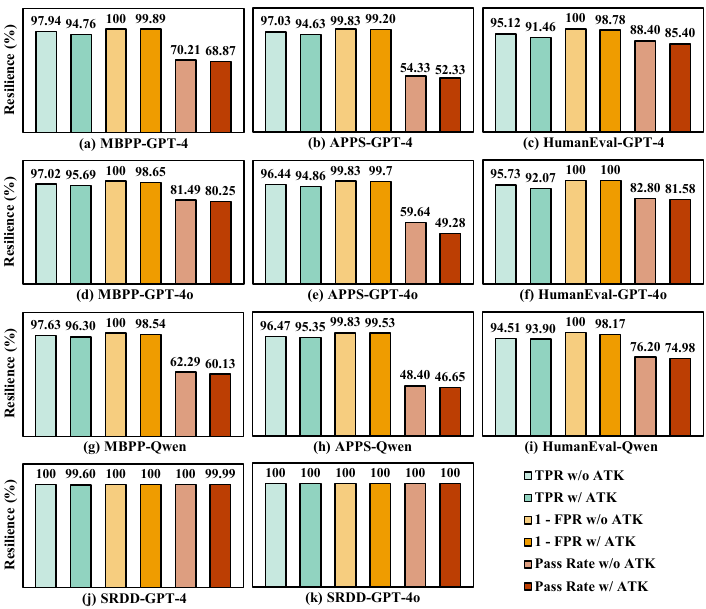}
    %\vspace{-10pt}
    \caption{Resilience results on content-focused code modifications.}
    \vspace{-5pt}
    \label{fig:attack2}
\end{figure}

Figure~\ref{fig:attack2} presents the resilience results under the modification attack.
As in the charts, \textbf{ACW} achieves promising TPR and FPR on all datasets without pass rate degradations.
For example, in Figure~\ref{fig:attack2} (a), \textbf{ACW} achieves TPR of 94.76\% and $1-$FPR of 99.89\% on MBPP-GPT-4, showing negligible degradations compared to the unattacked TPR of 97.94\% and $1-$FPR of 100\%, and retaining the pass rate of 70.21\% unchanged.
The promising results demonstrate the resilience of \textbf{ACW} against content modifications, as the detectability of our watermark is supported by a set of transformation rules, rather than relying on specific content in code.

We additionally explore the resilience of \textbf{ACW} in an extreme setting, where an attacker aims to entirely and arbitrarily rewrite the internal logic and control structure of code.
Our findings indicate that excessive modifications result in invalid attacks that are beyond real-world threat models, as the utility of code is significantly destroyed by the modifications.
In contrast, the above \emph{Attacker~2} provides a valid direction for reproducing the typical content modification attack on code.
As the attacker that violates validity prerequisites deviates from our evaluation objectives, we provide the supplementary results in an online appendix (as in Section~\ref{sec:availability}).

\emph{Attacker~3: Adaptive attacks.}
We assume that our watermark embedding knowledge in \textbf{ACW} can be leaked to the attacker, who aims to disrupt our watermark through adaptive attacks.
We implement the attacker by fine-tuning a CodeT5+~\cite{codet5plus}~(770M) LLM based on paired code sequences after and before watermarking.
The objective of the attacker LLM is to learn the code-to-code transformation that recovers a piece of watermarked code into its original version.

\begin{figure}[t]
    \centering
    \includegraphics[width=1\linewidth]{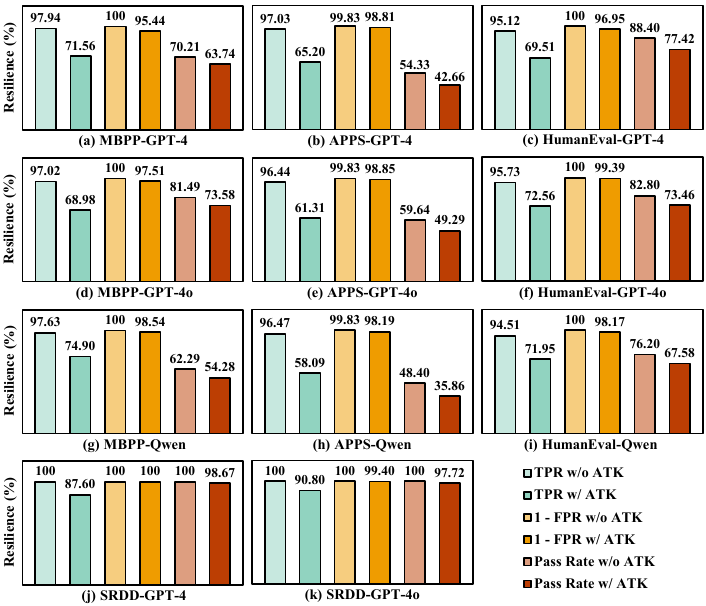}
    %\vspace{-10pt}
    \caption{Resilience results on adaptive attacks.}
    \vspace{-5pt}
    \label{fig:attack3}
\end{figure}

Figure~\ref{fig:attack3} presents the resilience results under the adaptive attack.
It is first observed that the attack is generally valid in terms of pass rate results, e.g., in Figure~\ref{fig:attack3} (a), the pass rate degrades from 70.21\% to 63.74\% on MBPP-GPT-4 after the attack, indicating that the utility of some code is destroyed, but the degree is within the scope of real-world threat models.
We further remark that the adaptive attack is forceful as the attacker LLM has learned the watermark embedding knowledge for recovering our watermarked code.
Under such premises, \textbf{ACW} is basically resilient against the adaptive attack with certain TPR degradations and stable FPR results, e.g., in Figure~\ref{fig:attack3} (a), the TPR is 71.56\% on MBPP-GPT-4 versus the unattacked TPR of 97.94\%.
The results indicate that despite multiple distinct transformations in code being destroyed by adaptive modifications, the remainders are still resilient to retain the discriminability.

\emph{Attacker~4: Watermark existence detection.}
The attacker aims to detect the existence of our watermark, thereby taking further disruptive actions as the above \emph{Attacker~1-3}.
In particular, we assume an adaptive attacker who is accessible to our watermark embedding knowledge and is able to craft both versions of code before and after watermarking.
The attacker then evaluates the invisibility of our watermark based on the functional semantic similarity (in terms of CodeBERTScore~\cite{zhou2023codebertscore}) and structural semantic similarity (in terms of CodeBLEU~\cite{ren2020codebleu}) between the original and watermarked code.
Note that CodeBERTScore and CodeBLEU are advanced metrics that measure code similarly comprehensively, providing forceful tools for the attacker to detect the existence of our watermark. 

Table~\ref{tab:invisibility} presents the invisibility results containing CodeBERTScore, CodeBLEU and their average scores, where the full score is 100.
For standard and fair evaluations, each CodeBERTScore and CodeBLEU result is measured as the average score of the total code in a dataset, with higher scores indicating better invisibility of our watermark.
As shown in the table, \textbf{ACW} achieves promising results with all average scores over 93, e.g., 93.73 on MBPP-GPT-4, indicating high similarity between the original and watermarked code.
The invisibility benefits from our semantic-preserving design for the transformation rules in \textbf{ACW}, making it challenging for the attacker to detect the existence of our watermark.

\emph{Based on this experiment, we conclude that \textbf{ACW} is resilient against potential code disruptions.}

\begin{table}[!t]
\caption{Evaluation Results on Invisibility}
\begin{center}
\scriptsize
\renewcommand\arraystretch{1.1}
\resizebox{0.48\textwidth}{!}
{\centering
\begin{tabular}{cccc}
\toprule
Dataset       
& CodeBERTScore  
& CodeBLEU  
& Average \\ 
\midrule 
MBPP-GPT-4               
& 99.10 & 88.35 & 93.73                            
\\
APPS-GPT-4               
& 98.80 & 90.96 & 94.88                            
\\
HumanEval-GPT-4          
& 99.04 & 89.44 & 94.24                            
\\
SRDD-GPT-4         
& 99.32 & 92.06 & 95.69                            
\\
MBPP-GPT-4o              
& 99.05 & 88.69 & 93.87                            
\\
APPS-GPT-4o              
& 99.27 & 92.99 & 96.13                            
\\
HumanEval-GPT-4o         
& 98.95 & 89.20 & 94.08                            
\\
SRDD-GPT-4o         
& 99.24 & 93.78 & 96.51                            
\\
MBPP-Qwen                
& 98.78 & 89.67 & 94.23                            
\\
APPS-Qwen                
& 98.07 & 89.67 & 93.87                            
\\
HumanEval-Qwen           
& 98.81 & 90.01 & 94.41                            
\\
\bottomrule 
\end{tabular}}
\end{center}
\label{tab:invisibility}
\vspace{-5pt}
\end{table}

\section{Ablation Study}
\label{sec:study}

In this section, we perform ablation studies to explore the effect of multiple factors on \textbf{ACW}, including (1) transformation idempotence, (2) transformation orders, (3) watermark overlaps and (4) code lengths.
Note that the following ablative results are counted based on all the function-level watermarked code in Section~\ref{subsec:setup}, as the project-level results are 100\% in different aspects (benefiting from abundant watermarking space), which produces opposite effects for analyzing the ablative results.

\subsection{Transformation Idempotence}
\label{subsec:idempotence}

We start with evaluating the idempotence of our transformations, i.e., applying a certain transformation once or multiple times results in the same code.

\begin{figure}[t]
    \centering
    \includegraphics[width=0.92\linewidth]{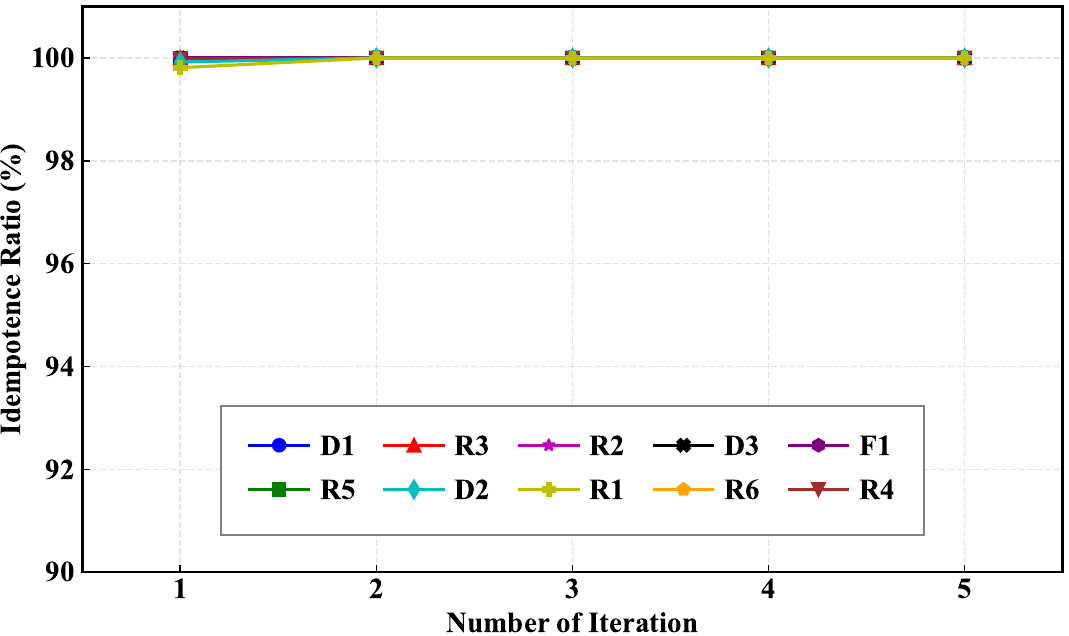}
    %\vspace{-10pt}
    \caption{Ablative results on the idempotence of transformations.}
    \vspace{-5pt}
    \label{fig:idempotence}
\end{figure}

Figure~\ref{fig:idempotence} presents our idempotence results.
As in the chart, we respectively apply the top-10 contributing transformation rules (as in Figure~\ref{fig:frequency}) over five iterations to the total watermarked code (in the horizontal axis), and count the idempotence ratio of each iteration (in the vertical axis), i.e., the proportion of code that remains unchanged under a given transformation among the total code.
It is shown by different colors of curves that only scarce watermarked code (less than 0.2\%) reveals changes under the first additional transformation, and remains unchanged in the subsequent iterations.
The promising results demonstrate that our designed transformations maintain perfect idempotence, ensuring the effectiveness of \textbf{ACW}.

\subsection{Transformation Order}
\label{subsec:order}

In the following, we evaluate whether the application order of transformation rules affects the effectiveness of our watermark.

Table~\ref{tab:order} presents our ablative results with different orders of transformations, including the average TPR and pass rates over the total watermarked code.
As in the table, three groups of code are watermarked based on different application orders of 46 transformation rules, where each order is generated using a distinct random seed.
It is shown by the results that three groups of transformations ordered by different random seeds achieve approximate TPR and pass rates (with differences below 0.2\%), demonstrating the stability and reliability of \textbf{ACW}.

\begin{table}[!t]
\caption{Ablative Results on Transformation Orders}
\begin{center}
\tiny
\resizebox{0.35\textwidth}{!}
{\centering
\begin{tabular}{ccc}
\toprule
Group    
& TPR (\%)  
& Pass Rate (\%)  \\ 
\midrule 
Random Seed~1               
& 97.03 & 59.08                            
\\
Random Seed~2             
& 96.98 & 59.10                            
\\
Random Seed~3         
& 97.10 & 59.11                             
\\
\bottomrule 
\end{tabular}}
\end{center}
\label{tab:order}
\vspace{-5pt}
\end{table}

\begin{table}[!t]
\caption{Ablative Results on Overlapped Watermarks}
\begin{center}
\Huge
\renewcommand\arraystretch{1.1}
\resizebox{0.48\textwidth}{!}
{\centering
\begin{tabular}{cccccc}
\toprule
Group
& \multicolumn{2}{c}{Embedding} 
& Identification 
& TPR (\%) 
& Fluctuation (\%) \\
\midrule
1  & \multirow{6}{*}{Seed 1~} & Set A & Set A & 97.36 & ---   \\
2  &                        & Set A$\rightarrow$Set B & Set A & 96.67 & 0.69 \\
3  &                        & Set A$\rightarrow$Set B$\rightarrow$Set C & Set A & 97.04 & 0.32 \\
4  &                        & Set A$\rightarrow$Set B & Set B & 97.56 & ---   \\
5  &                        & Set A$\rightarrow$Set B$\rightarrow$Set C & Set B & 97.78 & 0.22 \\
6  &                        & Set A$\rightarrow$Set B$\rightarrow$Set C & Set C & 98.03 & ---   \\
\midrule
7  & \multirow{6}{*}{Seed 2~} & Set D & Set D & 97.21 & ---   \\
8  &                        & Set D$\rightarrow$Set E & Set D & 97.03 & 0.18 \\
9  &                        & Set D$\rightarrow$Set E$\rightarrow$Set F & Set D & 97.19 & 0.02 \\
10 &                        & Set D$\rightarrow$Set E & Set E & 97.30 & ---   \\
11 &                        & Set D$\rightarrow$Set E$\rightarrow$Set F & Set E & 97.13 & 0.17 \\
12 &                        & Set D$\rightarrow$Set E$\rightarrow$Set F & Set F & 97.82 & ---   \\
\midrule
13 & \multirow{6}{*}{Seed 3~} & Set G & Set G & 97.11 & ---   \\
14 &                        & Set G$\rightarrow$Set H & Set G & 96.74 & 0.37 \\
15 &                        & Set G$\rightarrow$Set H$\rightarrow$Set I & Set G & 96.91 & 0.02 \\
16 &                        & Set G$\rightarrow$Set H & Set H & 97.31 & ---   \\
17 &                        & Set G$\rightarrow$Set H$\rightarrow$Set I & Set H & 97.79 & 0.48 \\
18 &                        & Set G$\rightarrow$Set H$\rightarrow$Set I & Set I & 97.93 & ---   \\
\bottomrule
\end{tabular}}
\end{center}
\label{tab:overlap}
\vspace{-5pt}
\end{table}

\subsection{Watermark Overlap}

We next evaluate whether applying our watermarks multiple times (with different transformation sets representing different LLMs) affects the effectiveness of \textbf{ACW}.

Table~\ref{tab:overlap} presents our evaluation results.
To take the first six experimental groups as an example, three transformation sets (Set~A, Set~B and Set~C), each containing 23 transformations, are selected using the random seed~1.
Note that the transformations in each set are randomly intersected rather than fully overlapping, as the latter has been evaluated in Section~\ref{subsec:idempotence}, where repeated transformations tend to produce the same code (as the idempotence results in Figure~\ref{fig:idempotence}).
In particular, we sequentially watermark our code using the transformations in Set A, Set B and Set C, producing overlapped watermarks.
In the table, Group 1 presents the initial TPR results by identifying the watermark in Set A.
Group 2-3 presents the TPR results and TPR fluctuations (compared to the initial TPR in Group 1) of Set A under sequential watermarks using Set B and Set C.
Similarly, Group 4-5 presents the initial and overlapped results of Set B, and Group 6 presents the final TPR results of Set C.
It is shown in the table that the TPR results remain stable (with TPR fluctuations below 1\%) when applying our transformations multiple times, which further demonstrates the discriminability and resilience of \textbf{ACW}.

\subsection{Code length}
\label{subsec:length}

In the end, we explore the effect of code lengths on the discriminability of our watermark.

\begin{figure}[t]
    \centering
    \includegraphics[width=1\linewidth]{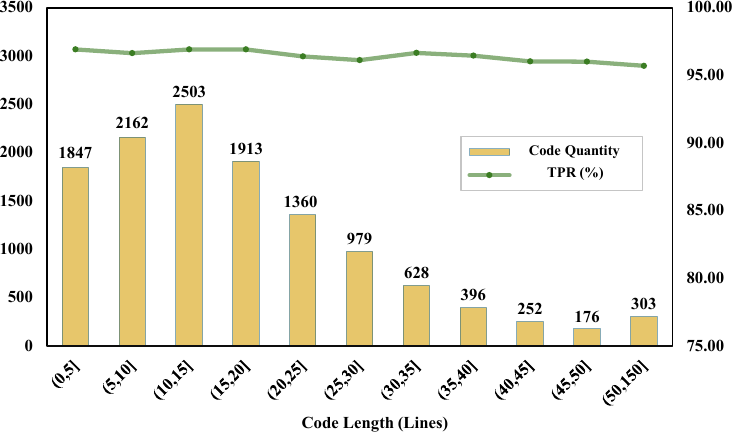}
    %\vspace{-10pt}
    \caption{Ablative results on the effect of code length on discriminability.}
    %\vspace{-5pt}
    \label{fig:ablation}
\end{figure}

Figure~\ref{fig:ablation} presents our ablative results.
As the horizontal axis in the chart, we group our watermarked code by different ranges of code length, where the yellow columns count the quantity of code in each group.
As the green curve in the chart, we then identify the watermark in each group and evaluate the discriminability in terms of TPR.
It is shown by the green curve that \textbf{ACW} presents stable TPR results over 95\% across all code lengths, which demonstrates our watermark can be successfully embedded and identified over different code lengths.

\section{Related Work}
\label{sec:relatedwork}

In this section, we review related work on AI-generated text and code detection, which is broadly related to this work.
In particular, existing approaches are commonly categorized into passive and active detectors.

\subsection{Passive Detection}

The passive approaches involve building binary classifiers to distinguish AI-generated or human-written content.
Currently, there is little work specifically focusing on AI-generated code, instead, they aim to detect any AI-generated content, regardless of text or code.
\citet{detect-gpt} show that AI-generated text tends to occupy the negative curvature region of LLM log probability functions, proposing DetectGPT as a detection tool. 
\citet{bao2024fast} introduce a conditional probability curvature to highlight discrepancies between AI and human text, presenting Fast-DetectGPT that improves the efficiency of DetectGPT. 
\citet{hu2023radar} report that AI detectors lack robustness against LLM-based paraphrasing and propose RADAR, an adversarial learning-based detection framework. 
\citet{gptzero} attempts to uncover the shortcomings of LLMs and propose a practical web application to detect AI-generated content.

While these passive detectors are effective at detecting AI-generated text, \citet{evaluating-aigc} report that existing approaches cannot generalize to code unless extensive additional fine-tuning is performed.
Additionally, similar to other domains like image forgery detection~\citep{imageforensics}, passive detectors are considered less generalizable than watermarking approaches, in which the embedded patterns are more stable than general features.

\subsection{Active Detection}

The active approaches involve embedding hidden watermarks into AI-generated content and detecting AI-generated or human-written content based on the presence of watermarks.

Similar to passive detectors, most of the LLM watermarking efforts focus on text.
As a pioneer approach, \citet{watermark-for-llm} propose WLLM, which splits a vocabulary into ‘green’ and ‘red’ tokens, prompts ‘green’ tokens during text generation, and identifies the presence of ‘green’ tokens to determine whether certain text is AI-generated. 
\citet{giboulot2024watermax} propose WaterMax, which generates multiple text chunks, selects the one that maximizes a predefined test statistic for watermarking, and detects watermarks by evaluating the test statistic.
\citet{zhaoprovable} propose Unigram-Watermark, which extends WLLM with a fixed token grouping strategy.
\citet{liu2024adaptive} propose AWTI, which performs adaptive watermark token identification, selectively targeting high-entropy tokens while leaving low-entropy tokens unchanged.
\citet{christ2024undetectable} propose a cryptographically-inspired undetectable watermark, detectable only with secret keys for preventing detectable changes to output distributions. 
\citet{zhang2024remark} propose REMARK-LLM, a watermarking framework comprising an encoding module for embedding binary signatures, a reparameterization module for mapping signatures into sparse tokens, and a decoding module for watermark extraction.   
\citet{dathathri2024scalable} propose SynthID-Text, an LLM text watermarking scheme that operates by modifying the sampling procedure during text generation.

Compared to text watermarking, detecting AI-generated code based on watermarking remains an emerging task, with limited research to date.
Based on the red-green watermarking strategy from WLLM~\citep{watermark-for-llm}, \citet{who-wrote} propose a selective watermarking technique via entropy thresholding called SWEET, which extends WLLM by prompting ‘green’ tokens only at high-entropy positions, so as to preserve code utility.
\citet{kim2025marking} further extend SWEET and propose a syntax token-preserving watermarking approach, called STONE, which excludes tokens critical to code execution, thereby avoiding utility degradations.

Based on our experimental evaluations in Section~\ref{sec:experiment}, existing state-of-the-art approaches demonstrate limited accuracy in detecting LLM-generated code, and present significant time and memory consumption that slows down LLMs.

In addition to the watermarking approaches, some research explores attacking strategies for benchmarking the resilience or robustness of LLM watermarks, and existing efforts mainly focus on AI-generated text.
For example, focusing on the red-green watermarking strategy as in WLLM~\citep{watermark-for-llm}, \citet{zhang2024stealing} propose a ‘green’ list stealing attack, which formalizes a mixed integer programming problem, with the aim to find a minimal available ‘green’ list constrained by a set of rules.
\citet{pang2024no,pang2024attacking} explore the trade-off in the robustness, utility and usability of LLM watermarks, and propose a content-based spoofing attack that results in the watermarked text being inaccurate through content modifications.
Based on our experimental evaluations in Section~\ref{sec:experiment}, \textbf{ACW} is resilient
against multiple types of typical attacks.

Apart from watermarking approaches for AI-generated code detection, some research explores code watermarking in other applications.
\citet{guan2024codeip} extend WLLM and propose CodeIP, which embeds information with crucial provenance details for safeguarding the intellectual property of LLMs. 
\citet{ning2024mcgmark} also extend WLLM and propose MCGMark, which selects tokens based on probabilistic outliers to trace malicious code.
\citet{llm-ip} introduce ToSyn, a watermarking scheme that stealthily tweaks the distribution among synonym tokens in LLM outputs, for protecting LLM APIs from remote imitation attacks. 
\citet{codemark} propose CodeMark, which embeds user-defined imperceptible watermarks into code datasets to trace their usage in code completion models. 
\citet{yang2024srcmarker} introduce SrcMarker, which uses learning-based embedding and extraction strategies to select rules for watermarking, focusing on traditional source code watermarking unrelated to LLMs. 
These methods explore diverse applications of code watermarking, beyond the task of AI-generated code detection.

\section{Conclusion and Discussion}
\label{sec:conclusion}

In this work, we propose a plug-and-play code watermarking approach, i.e., \textbf{ACW}, for AI-generated code detection.
\textbf{ACW} watermarks AI-generated code based on a series of carefully-designed, semantic-preserving and idempotent code transformations.
\textbf{ACW} detects AI-generated code by identifying the existence of our watermarks, i.e., whether our transformations have been applied.
Experimental results demonstrate that \textbf{ACW} is effective in detecting AI-generated code, preserves code utility, and is resilient against code optimizations.
Especially, \textbf{ACW} is efficient and universal across different LLMs, addressing the limitations in existing work.

Finally, we discuss potential validity threats regarding \textbf{ACW}. 
First, \textbf{ACW} currently includes three groups of 46 transformations, which are demonstrated effective in watermarking and identifying AI-generated Python code.
It is straightforward that our rule library can be naturally expanded from both aspects of rule numbers and other languages, based on rich language features provided by modern programming languages.
Second, \textbf{ACW} focuses on an emerging task driven by the development of LLMs, i.e., AI-generated code detection.
In the future, our watermarking strategies have the potential to extend to broader applications, e.g., tracing the authorship of open-source code or projects, verifying the integrity of code over time, and identifying reused code in copyright-sensitive projects.

\section{Code and Data Availability}
\label{sec:availability}

We release our code and data resulting from this work, along with an online appendix presenting our complete transformation rules and supplementary experiments, available at~\url{https://github.com/Noelle1831-k/ACW}.

\section*{Acknowledgment}

This work is supported by the SMU-SUTD Internal Research Grant (code: MSS24C010).

We sincerely appreciate the editors and reviewers for their insightful feedback and constructive suggestions throughout the review process, which have contributed to the improvement and refinement of this paper.

\bibliographystyle{IEEEtranN}
\bibliography{IEEEabrv,reference}

\vfill

\end{document}